\documentclass[11pt,a4paper]{article}

\usepackage[implicit=false]{hyperref}
\usepackage[mathscr]{euscript}
\usepackage[textsize=scriptsize,textwidth=2.5cm]{todonotes}
\usepackage{amssymb,amsmath,latexsym,bm,amsfonts,mathtools}
\usepackage{graphicx}
\usepackage{longtable}
\usepackage{color}
\usepackage{indentfirst}
\usepackage{subfig}
\usepackage{comment}
\usepackage[normalem]{ulem}
\usepackage{array}
\usepackage{tikz-cd}
\tikzcdset{scale cd/.style={every label/.append style={scale=#1},
    cells={nodes={scale=#1}}}}

\usepackage{caption} 
\usepackage{floatrow}

\usepackage{multirow}
\usepackage{booktabs}
\usetikzlibrary{positioning, shapes.geometric, patterns, graphs, decorations.pathmorphing}
\usepackage{relsize}
\usepackage{graphicx}
\usepackage{calc}

\numberwithin{equation}{section}

\newcommand{\N}{\ensuremath{\mathbb{N}}}
\newcommand{\Z}{\ensuremath{\mathbb{Z}}}

\newcommand{\R}{\ensuremath{\mathbb{R}}}
\newcommand{\CC}{\ensuremath{\mathbb{C}}}

\def \x#1 {x_{#1}}

\newcommand{\bra}{\left\langle}
\newcommand{\ket}{\right\rangle}

\newcommand{\bO}{\boldsymbol{O}}
\newcommand{\bT}{\boldsymbol{T}}
\newcommand{\bxi}{\boldsymbol{\xi}}
\def \D {\Delta}
\newcommand{\eq}[1]{\begin{align}#1\end{align}}

\def \p {\partial}
\def \ba {\begin{array}}
\def \ea {\end{array}}

\definecolor{oliveyuan}{RGB}{154,205,50}

\begin{document}

\title{A BMS-invariant free scalar model}

\author{
Peng-xiang Hao$^{a}$,
Wei Song$^{b,c}$,
Xianjin Xie$^a$,
and Yuan Zhong$^a$\footnote{pxhao@mail.tsinghua.edu.cn, wsong2014@mail.tsinghua.edu.cn, xxj19@mails.tsinghua.edu.cn, zhongy17@mails.tsinghua.edu.cn.}
}
\date{}

\maketitle

\begin{center}
{\it
$^{a}$Yau Mathematical Sciences Center, Tsinghua University, Beijing 100084, China\\
\vspace{2mm}
$^{b}$Department of Mathematical Sciences, Tsinghua University, Beijing 100084, China\\
$^{c}$Peng Huanwu Center for Fundamental Theory, Hefei, Anhui 230026, China}
\vspace{10mm}
\end{center}

\begin{abstract}

The BMS (Bondi-van der Burg-Metzner-Sachs) symmetry arises as the asymptotic symmetry of flat spacetime at null infinity. In particular, the BMS algebra for three dimensional flat spacetime (BMS$_3$) is generated by the super-rotation generators which form a Virasoro sub-algebra with central charge $c_L$, together with mutually-commuting super-translation generators. The super-rotation and super-translation generators have non-trivial commutation relations with another central charge $c_M$.
In this paper, we study a free scalar theory in two dimensions exhibiting BMS$_3$ symmetry, which can also be understood as the ultra-relativistic limit of a free scalar CFT$_2$ in the flipped representation. Upon canonical quantization on the highest weight vacuum, the central charges are found to be $c_L=2$ and $c_M=0$. Because of the vanishing central charge $c_M=0$,  the theory features novel properties: there exist primary states which form a multiplet, and the Hilbert space can be organized by an enlarged version of BMS modules dubbed the staggered modules.
We further calculate correlation functions and the torus partition function, the latter of which is also shown explicitly to be modular invariant.  Is it interesting to note that our model has vanishing $c_M$, a feature also shared by the so-called flat space chiral gravity in \cite{Bagchi:2012yk}.

\end{abstract}

\baselineskip 18pt
\thispagestyle{empty}

\newpage
\tableofcontents
\newpage

\section{Introduction}
As the first example of unified space and time, Minkowski spacetime is in some sense the starting point of both quantum field theory and general relativity:
it is usually the spacetime background for Lorentzian invariant quantum field theories, and it also provides the simplest solution to Einstein equation. Yet it is still a mystery what the full-fledged quantum theory of gravity on asymptotically Minkowski spacetime should be.

In quest of quantum gravity, one fruitful effort over the past  two decades  has been holographic duality, whose most well-known  incarnation is the  AdS/CFT correspondence \cite{Maldacena, Gubser-Klebanov-Polyakov, Witten}.  The AdS/CFT correspondence  states that gravity in asymptotically Anti-de Sitter spacetime is equivalent to quantum field theory with conformal invariance, and it has  become a thriving research field involving many interdisciplinary studies including string theory, black hole physics, condensed matter physics, quantum chromodynamics, and quantum information theory.

It is a tantalizing idea to extend the success of the AdS/CFT correspondence to asymptotically flat spacetime.
To this end, asymptotic symmetries play an important role.
Recall that in AdS$_{d+1}$/CFT$_d$, one basic item in the holographic dictionary is that the asymptotic symmetry of the bulk theory agrees with the global symmetry of the dual field theory, and both are the conformal symmetry in $d$ dimensions. For instance, under the Brown-Henneaux boundary conditions \cite{Brown-Henneaux}, the asymptotic symmetry for Einstein gravity with negative cosmological constant $-\frac{1}{\ell^2}$ agrees with that of a CFT$_2$ with central charges $c_L=c_R=\frac{3\ell}{2G}$, where $G$ is Newton's constant in three dimensions. The symmetry argument is especially powerful in the case of AdS$_3$/CFT$_2$, using which one can provide a microscopic explanation of the Bekenstein-Hawking entropy of the black holes using Cardy's formula \cite{Strominger-Vafa, Strominger}.
The asymptotic symmetry for four-dimensional Minkowski spacetime in Einstein gravity, first studied by Bondi-van der Burg-Metzner-Sachs (BMS) \cite{Bondi-van der Burg-Metzner, R. K. Sachs}, is the so-called BMS symmetry. The original BMS symmetry only contains generators that are smooth on $S^2$.
In a recent resurgence \cite{Hawking:2016msc,
Hawking:2016sgy, Strominger:2013lka, Strominger review}, the extended version of BMS group also admits generators that are singular at the south or north poles. The extended BMS symmetry in four dimensions is related to Weinberg's soft theorem and the memory effect \cite{Strominger:2014pwa}, and has recently prompted the study of celestial CFTs, see \cite{Raclariu:2021zjz, Pasterski:2021rjz, Aneesh:2021uzk}.

A simpler version of the BMS group appears in three dimensions \cite{Ashtekar:1996cd, Compere-Barnich,Barnich:2012aw}.
Under certain boundary conditions at null infinity, the asymptotic symmetry for flat spacetime in Einstein gravity is generated by the superrotations $L_n$ and supertranslations $M_n$, where $n$ can be arbitrary integers. The BMS algebra is,
\eq{\label{BMSalgintro}
\left[L_{n},\,L_{m}\right] &= (n-m)L_{m+n}+\frac{c_L}{12} n(n^2-1)\delta_{m+n,0},\nonumber\\
	\left[L_{n},\,M_{m}\right] &= (n-m)M_{m+n}+\frac{c_M}{12}  n(n^2-1)\delta_{m+n,0},\\
	 \left[M_{n},\,M_{m}\right] &= 0\nonumber\,.
}
The superrotation generators form a Virasoro subgroup, with central charge denoted by $c_L$.
The supertranslation generators $M_n$ commute with each other, but have a non-trivial commutation relation with the Virasoro generators with central charge $c_M$.
 The BMS$_3$ algebra \eqref{BMSalgintro} is isomorphic to the Galilean conformal algebra (GCA) in two dimensions \cite{2010,2009,Bagchi:2012cy}. While the GCA can be obtained from the non-relativistic (NR) limit of the two-dimensional conformal algebra,
 the BMS$_3$ algebra is the ultra-relativistic (UR) limit, and thus is also an example of a Carrollian algebra \cite{carrollian,BinChen01}.
 Like their CFT$_2$ cousins, field theories invariant under BMS or Galilean conformal symmetries are highly constrained.  In particular, symmetry and other consistency conditions make it possible to initiate a bootstrap program \cite{bootstrap, Bagchi:2016geg, Bagchi:2017cpu, BinChen02, BinChen03,Chen:2019hbj}.

 It is reasonable to conjecture that the holographic dual of Einstein gravity in asymptotically flat three-dimensional spacetimes is a quantum field theory invariant under the BMS$_3$ symmetry (BMSFT). As evidence, the torus partition function for BMSFTs has been argued to be modular invariant and a Cardy-like formula can be used to explain the entropy of cosmological solutions with Cauchy horizons \cite{Bagchi:2012xr, Barnich:2012xq}.
 Furthermore, entanglement entropy and its holographic dual has been calculated in \cite{Bagchi:2014iea, wilson line paper,  Song-Wen-Jiang, swing paper, Apolo:2020qjm}.
 Other interesting properties in flat holography include geometric Witten diagrams \cite{Hijano:2017eii}, quantum energy conditions \cite{Grumiller:2019xna}, etc.

Despite this progress, we know little about the putative dual field theory other than properties that can be implied by the symmetries.
In particular, it is important to know if a field theory with BMS invariance really exists at the full quantum level.
Thus, it is necessary to construct and study in detail an explicit model of BMSFT.
Besides the motivation from flat holography, a model of BMSFT is also interesting from a purely field theoretic perspective, as it provides a playground for further understanding both non-relativistic  and ultra-relativistic quantum systems.
A Liouville-like theory with BMS symmetries has been constructed in \cite{Barnich:2012rz,Barnich:2013yka}, which can be obtained from the ultra-relativistic limit of ordinary Liouville theory, or alternatively from the geometric action for the BMS$_3$ group \cite{Barnich:2017jgw}.

In this paper, we study a free scalar BMSFT model in two dimensions with the action
\eq{
S = \frac{1}{4\pi} \int d\sigma d\tau \left( \p_{\tau} \phi \right)^2. \label{actionintro}
}
The classical theory is invariant under the BMS symmetry \eqref{BMSalgintro} with zero central charges.
This model also appears in the tensionless limit of string theory \cite{Bagchi}, and a $\sqrt{T\bar T}$ deformation of a free scalar CFT$_2$ \cite{Troncoso}.

After canonical quantization and a choice of the vacuum compatible with the highest weight representation, the BMS algebra of the model \eqref{actionintro} has central charges \eq{\label{cc13} c_L=2, \quad c_M=0.}
Due to the non-trivial commutation relations between $L_m$ and $M_n$, the action of $M_0$ is not necessarily diagonal, and there can exist multiplets on which the action of $M_0$ is a Jordan cell with all the diagonal components equal and denoted by $\xi$.
Multiplets are thus labeled by the conformal weight $\Delta$, which is the eigenvalue of $L_0$, and the boost charge $\xi$ which comes from the Jordan cell of $M_0$.
The model \eqref{actionintro} has the following key features,
\begin{itemize}
\item The fundamental primary operators are \eq{I, \quad  \bO=(O_0=i\p_y \phi, \, O_1=i\p_x \phi), \quad V_\alpha=:e^{\alpha \phi}:,}  where $I$ is the identity operator, $\bO$ is a primary multiplet with $\Delta=1,\, \xi=0$, and $V_\alpha$ is the vertex operator with $\Delta=0,\, \xi=-{\alpha^2\over2}$. Correlation functions between these operators can be calculated explicitly.
\item States are organized in terms of an enlarged version of BMS highest weight module dubbed the staggered BMS module.
\item The torus partition function can be calculated and is found to be modular invariant, confirming earlier statements based on symmetry arguments \cite{Bagchi:2012xr, Barnich:2012xq, Song-Wen-Jiang, Bagchi:2019unf}.
\end{itemize}

The appearance of primary multiplets and staggered modules are both unexpected. In \cite{2010, bootstrap},  it has been noticed that the commutation relations \eqref{BMSalgintro} between $M_0$ and $L_n$ imply that the action of $M_0$ is not diagonal within a BMS highest weight module, and descendants have to form multiplets. In particular, the current $T$ which generates the superrotations, and the current $M$ which generates the supertranslations form a multiplet with conformal weight $\Delta=2$ and boost charge $\xi=0$.
While the algebra \eqref{BMSalgintro} implies that multiplets at the level of BMS descendants are inevitable,
there is no {\it a  priori} reason that multiplets for primary operators have to exist as well. Our model thus provides the first example of this novel representation.

The appearance of the staggered module is closely related to the subtlety with $c_M=0$, at which point it was argued that the BMS highest weight module is truncated to the Virasoro module in \cite{2010}. Instead of a truncation, however, here the BMS highest weight module is enlarged. The reason for this is that there is an extra quasi-primary $K$ with conformal weight $\Delta=2$ in our model that was not assumed to exist in the general argument \cite{2010}. The new quasi-primary $K$ forms a BMS triplet together with the Virasoro stress tensor $T$ and supertranslation stress tensor $M$. This structure is reminiscent of logarithmic CFTs \cite{Gaberdiel:1996kx, Rohsiepe:1996qj, Gaberdiel:2001tr, Kytola:2009ax, Cardy:2013rqg, Creutzig:2013hma}, where the Virasoro stress tensor acquires a logarithmic partner whose presence makes the action of $L_0$ non-diagonal. In that case, multiplets also appear and the states are also organized into staggered modules \cite{Gaberdiel:1996kx, Kytola:2009ax}.
It would be interesting to further understand the staggered module from a general analysis, and also to study the implications to holography.
We will leave these questions for further study.

Apart from the intrinsic discussion as a two dimensional BMSFT, the model \eqref{actionintro} with the central charges \eqref{cc13} can also be obtained from a free scalar CFT$_2$ by two steps.
From a free scalar CFT$_2$ with $c=1,\bar c=1$, we can construct the so-called flipped representation of the conformal algebra with central charges $c=1,\bar c=-1$. Then our BMSFT model can be obtained from the UR limit of the CFT$_2$ in the flipped representation. Note that this flippling $+$ UR limit is different from the direct UR limit discussed in \cite{Barnich:2012aw,Barnich:2013yka}.

Finally, let us briefly comment on the potential gravity dual to this model. It is not directly applicable to Einstein gravity, as the latter has the central charges $c_L=0, \, c_M={3\over G}$. To find BMS$_3$ with non-vanishing $c_L$, \cite{Hotta:2010qi, Bagchi:2010vw} considered topologically massive gravity (TMG) which is three dimensional Einstein gravity plus a gravitational Chern-Simons \cite{Deser:1981wh,Deser:1982vy}. Let the coupling in the  Chern-Simons term $\mu$, the central charges are given by \eq{c_L=\frac{3}{\mu G},\ \ c_M=\frac{3}{G}.}
To get vanishing $c_M$, we need to take the limit where only the Chern-Simons term persists, which is the so-called flat space chiral gravity theory as discussed in \cite{Bagchi:2012yk}.

This paper is organized as follows. Section 2 is a general analysis of BMSFTs. In section 2.1, we first briefly review the general properties of BMSFTs,
in section 2.2 we discuss the representations, and introduce a novel highest weight representation where the action of $M_0$ is a Jordan cell,
and in section 2.3 we calculate the correlators. In section 3, we study the free BMS scalar model in detail.  We introduce the classical theory and write down its symmetries in section 3.1, perform a canonical quantization with the highest weight vacuum in section 3.2, and calculate correlation functions in section 3.3.
In section 4, we arrive at the key result of this paper, the staggered BMS module. We find the operators outside the ordinary BMS highest weight module, and organize them into the staggered module. We further illustrate the properties of this module by diagrams. In section 5, we review the UR limit from two-dimensional CFTs to BMSFTs, and point out a subtlety on the consistent plane UR limit. Then we take the UR limit of the free relativistic scalar model in the flipped vacuum with the central charges $c=1,\bar{c}=-1$ to get the free BMS scalar in the highest weight representation. In section 6, we calculate the torus partition function of the free BMS scalar model in the highest-weight representations, and find that it is invariant under modular S-transformations.

\section{General properties of BMSFTs}
In this section we discuss generic features of BMSFTs. We first provide a short review of the BMS algebra in section 2.1.
Section 2.2 is dedicated to the representation theory where we introduce a novel type of highest-weight representations, where the primary states lie in multiplets.  In section 2.3 we calculate the correlation functions for general quasi-primary multiplets, paying special attention to multiplets with $\xi=0$ which we will encounter later in the free scalar model.

\subsection{Quick Review}

A BMSFT (BMS-invariant field theory) is a two dimensional quantum field theory invariant under the following BMS transformation,
\begin{equation}
\sigma\to f(\sigma), \quad \tau\to f^{\prime}(\sigma)\tau + g(\sigma).\label{BMSfinite}
\end{equation}
Note that although the theory is not Lorentz invariant, there is still a notion of time and space. For BMSFT, $\sigma$ should be regarded as a spatial direction, whereas $\tau$ is a timelike direction. This interpretation will become clearer if we
view BMSFT as the ultra-relativistic (UR) limit of a CFT$_2$, to be described momentarily.

Now let us consider a BMSFT on a cylinder parameterized by the coordinates $(\tau, \,\sigma)$ with the identification \eq{\sigma\sim \sigma+2\pi.\label{sigmacircle}} Then the infinitesimal BMS transformation is generated by the Fourier modes,
\begin{align}
l_{n} &= ie^{in\sigma }\partial_{\sigma} -n e^{in\sigma}\tau\partial_{\tau}\\
m_{n} &= ie^{in\sigma }\partial_{\tau}\label{lmncylinder}
\end{align}
Under the Lie bracket,  the generators \eqref{lmncylinder}  form the BMS algebra
\eq{
\left[l_{n},l_{m}\right] &= (n-m)l_{m+n}\\
	\left[l_{n},m_{m}\right] &= (n-m)m_{m+n},\\
	 \left[m_{n},m_{m}\right] &= 0.
}
The generators that implement the transformations \eqref{lmncylinder} on the fields will be denoted as $L_m$ and $M_n$, and they
form the centrally extended BMS algebra,
\eq{\label{BMSalg}
\left[L_{n},\,L_{m}\right] &= (n-m)L_{m+n}+\frac{c_L}{12} n(n^2-1)\delta_{m+n,0}\nonumber,\\
	\left[L_{n},\,M_{m}\right] &= (n-m)M_{m+n}+\frac{c_M}{12}  n(n^2-1)\delta_{m+n,0}\nonumber,\\
	 \left[M_{n},\,M_{m}\right] &= 0,
}
where $c_L$ and $c_M$ are central charges. As a side remark, it is found that the asymptotic symmetry of Einstein gravity in asymptotically-flat three-dimensional spacetime is BMS$_3$ with $c_L=0,\,c_M={3 \over G}$ \cite{Compere-Barnich}, while  gravitational theories where both central charges do not vanish can be constructed by adding a Chern-Simons term \cite{Bagchi:2012yk,Hotta:2010qi,Bagchi:2010vw}.

The general form of the BMS transformation \eqref{BMSfinite} allows the map
\begin{equation}x=e^{i\sigma},\ \ \ y=i\tau e^{i\sigma} \label{plane2cylinder}\end{equation}
By analytic continuation, $x$ can be viewed as a holomorphic coordinate on the plane.  The above map \eqref{plane2cylinder} is usually regarded as the map from the cylinder to the plane \cite{Bagchi}, using which we can further discuss the state-operator correspondence.
For later convenience, we also write down the BMS generators on the plane,
\begin{align}
l_{n} &= -x^{n+1}\partial_{x} - (n+1)yx^{n}\partial_{y},\\
m_{n} &= -x^{n+1}\partial_{y}.
\end{align}

Let $T$ be the Noether current of the translational symmetry along $x$, and $M$ be the Noether current of the translational symmetry along $y$. The BMS charges on the plane can then be written as
\eq{
L_n&={1\over 2\pi i}\oint \Big( x^{n+1} T+  (n+1)x^{n}y M \Big),\\
M_n&= {1\over 2\pi i}\oint x^{n+1} M,}
where $\oint$ denotes the contour integration around the origin on the complexified $x$ plane.
\eq{
T&=\sum\limits_{n} L_n x^{-n-2}-\sum\limits_{n} (n+1)yM_{n-1}x^{-n-2},	\label{T_modes}\\
M&=\sum\limits_{n} M_n x^{-n-2}.	\label{M_modes}
}
From the algebra \eqref{BMSalg} we expect the following OPEs between the currents,
\eq{\label{BMSope}
T(x',y') T(x,y) & \sim \frac{c_L}{2(x'-x)^4}+ \frac{2T(x,y)}{(x'-x)^2}+\frac{\partial_x T(x,y)}{x'-x}  \\
&-\frac{2c_M(y'-y)}{(x'-x)^5}- \frac{4(y'-y) M(x,y)}{(x'-x)^3}- \frac{(y'-y)\partial_y T(x,y)}{(x'-x)^2}, \nonumber\\
T(x',y') M(x,y) & \sim {c_M\over 2 (x'-x)^4}+\frac{2M(x,y)}{(x'-x)^2}+\frac{\partial_x M(x,y)}{x'-x},\nonumber\\
M(x',y') M(x,y) & \sim 0.\nonumber
}
The transformation laws of the currents under the BMS transformation
\eq{\tilde{x}= f(x), \quad \tilde{y}= f^{\prime}(x)y + g(x)\label{BMSplane}
}
are given by \footnote{The  transformation law \eqref{translaw} is consistent with the OPE \eqref{BMSope} and successive transformations. \eqref{translaw} is also compatible with \cite{Song-Wen-Jiang}, after we match the conventions $J^{\text{there}}=T^{\text{here}}-y\p_x M^{\text{here}}$. We also note that the results of \cite{wilson line paper} differ from ours in the term proportional to $c_M$. For the transformation law, please see also \cite{Barnich:2012rz}. }
\eq{\label{translaw}
\tilde{M}(x)&=f^{\prime 2} M\left(\tilde{x} \right)+\frac{c_{M}}{12}\{f, x\},
\\\tilde{T}(x, y)&=f^{\prime 2} T\left(\tilde{x}, \tilde{y}\right)+2 f^{\prime } \big(g^{\prime}+y f''\big) M\left(\tilde{x}\right)+\frac{c_{L}}{12}\{f, x\}+ \frac{c_{M}}{12} \big(y{d\over dx}\{f,\,x\}+ f'^2 {\p^3 g\over \p f^3}\big).\nonumber
}
 In Eq.\eqref{translaw}, $\{\,,\,\}$ denotes the usual Schwarzian derivative, and the last term is the so-called BMS Schwarzian derivative \cite{wilson line paper, Song-Wen-Jiang},
\eq{
\{f,\,x\}&=\frac{{f}^{\prime \prime \prime}}{{f}^{\prime}}-\frac{3}{2}\left(\frac{{f}^{\prime \prime}}{{f}^{\prime}}\right)^{2},\\
f'^2 {\p^3 g\over \p f^3} &=f^{\prime-1}\left(g^{\prime\prime\prime}-g^{\prime} \frac{{f}^{\prime \prime \prime}}{{f}^{\prime}}-3 f^{\prime \prime}\left(\frac{g^{\prime}}{f^{\prime}}\right)^{\prime}\right).
}

\subsection{Representations}
In this subsection we discuss the representations of the BMS algebra \eqref{BMSalg}.
In the literature, two types of representations have been discussed, the highest weight representation and the induced representation \cite{Bagchi:2009ca}.
The induced representation is unitary and can be obtained from a Ultra relativistic limit of the highest weight representation of CFT$_2$s. The highest weight representation of BMS algebra is analogous to that of Virasoro, which enables one to adapt techniques of CFT$_2$ for BMSFT. Despite of the fact that the highest weight representation is non-unitary, several interesting results have been worked out in the highest weight representation including the general structure of correlation function, characters, torus partition function, entanglement entropy, bootstrap, and so on. We will focus on the highest weight representation in this section, and leave the discussion of induced representation to Appendix B.
In the following, we first review the usual highest weight representation discussed previously in the literature \cite{Bagchi:2009ca}, which we refer to as the singlet version. Then we will introduce a novel multiplet version of the highest weight representation. In section 3 we will see that the multiplet version representation arises naturally in the free scalar model \eqref{actionintro}.

In this subsection we discuss the highest weight representations of the BMS algebra \eqref{BMSalg}, which are not unitrary. We first review the usual highest weight representation discussed previously in the literature \cite{Bagchi:2009ca}, which we refer to as the singlet version. Then we will introduce a novel multiplet version of the highest weight representation. In section 3 we will see that the multiplet version representation arises naturally in the free scalar model \eqref{actionintro}.

\subsubsection{Highest weight representations: the singlet}

The singlet version of the highest weight representation of the BMS algebra \cite{Bagchi:2009ca} is a straightforward generalization of the highest weight representation of the Virasoro algebra.
This amounts to considering the BMS module on the plane which consists of primary operators and their descendants.
A primary operators at the origin $O=O(0,0)$ can be labelled by the eigenvalues $(\Delta,\xi)$ of $(L_0,M_0)$
\begin{equation}
[L_0,O]=\Delta O,\qquad [M_0,O]=\xi O.\label{eigen}
\end{equation}
$\Delta$ and $\xi$ are referred to as the conformal weight and the boost charge of the operator respectively. The highest weight conditions are
\begin{equation}
[L_n,O]=0,\ \ \ \ [M_n,O]=0, \ \ \ n>0.\label{highest}
\end{equation}
The descendant operators can be obtained by acting $L_{-n},M_{-n}$ with $n>0$ successively on the primary operators.  The primary operator together with its descendants form a highest weight module.

\subsubsection{Highest weight representations: the multiplet}
In any unitary theory, if two Hermitian operators commute, then we can go to a basis in which the commuting operators are simultaneously diagonalized.
Therefore it is natural to consider highest weight representations in CFT$_2$, and organize states into Virasoro modules.
For BMSFT, however, several subtleties arise in the highest weight representation of the form \eqref{eigen} and \eqref{highest}.
As noticed in \cite{2010}, the Kac determinant for the highest weight representation with $\xi\neq0$ is negative, and hence the representation is not unitary. In a state space equipped with an indefinite inner product, a Hermitian operator is not necessarily diagonalizable. In addition, it has been observed in \cite{Bagchi:2009ca} that if we organize the highest weight module in terms of quasi-primaries and their global descendants,
the quasi-primaries will generically form multiplets, on which the action of $L_0$ and $M_0$ cannot be simultaneously diagonalized within a BMS module even assuming the primary state is a common eigenstate of $L_0$ and $M_0$.  These observations then open up the possibility that the action of $L_0$ and $M_0$ is not diagonal even on the primary states. This feature is very similar to logarithmic CFTs \cite{Gaberdiel:1996kx, Rohsiepe:1996qj, Gaberdiel:2001tr, Kytola:2009ax, Cardy:2013rqg, Creutzig:2013hma}.
In this case, the representation matrix can be written in the Jordan canonical form. Similar to the discussion in \cite{Gaberdiel:2001tr}, we can choose a basis so that the action of $L_0$ is diagonal and the action of $M_0$ is block diagonal, with each block being a Jordan cell. The primary operators in a Jordan chain form a multiplet, which, together with their descendants, form a reducible but indecomposable module.
If there are $r$ operators related to each other in a Jordan chain, the multiplet they form will be referred to as having rank $r$ , the same rank as the Jordan block. The primary operators with diagonal action under $M_0$ will be referred to as singlets or rank-$1$ multiplets.

Thus operators of BMSFT can  be organized into highest-weight primary multiplets and their descendants. 
A highest-weight primary multiplet $\bO$ with rank $r$ is defined by
\eq{
[L_0,\,O_{a}]&=\Delta O_{a},\qquad [M_0,O_{a}]=({{\bxi}} O)_a,\quad a=0,\cdots r-1\nonumber\\
[L_n,\,O_{a}]&=0,\quad  [M_n,O_{a}]=0,  \quad n>0
,\label{multiplet}
}
where $O_{a}$ denotes the $a$-th component of the multiplet $\bO$, and ${{\bxi}}$ is a Jordan cell with rank $r$ and diagonal component $\xi$,
\begin{equation}\label{jordan}
\bxi=
\begin{pmatrix}
 \xi& & & \\
  1& \xi& & \\
  & \ddots&\ddots &\\
   & & 1& \xi\\
\end{pmatrix}_{r\times r}.
\end{equation}
The off-diagonal element can also be chosen to be any arbitrary constant, which amounts to introducing a relative scaling among different components.

Similar to CFT$_2$, it is also useful to introduce the notion of quasi-primary multiplets, which satisfy \eqref{multiplet} but only with $n=1$ rather than arbitrary positive integers.
Quasi-primary multiplets are highest weight states under the action of the global subgroup of the BMS$_3$ group, which is isomorphic to the Poincar\'e group in  three dimensions.

\subsection{Correlation functions}
Discussions on the correlation functions for quasi-primary operators of the Galilean conformal algebra can be found in \cite{2010, Chen:2019hbj}. The results in \cite{2010} directly apply to BMSFT, as the algebras of GCFT and BMSFT are isomorphic.  Interestingly, it has been found that multiplets appear generically in GCFTs and BMSFTs \cite{bootstrap}, a feature also shared by logarithmic CFTs \cite{Gaberdiel:1996kx, Rohsiepe:1996qj, Creutzig:2013hma}. Detailed discussions of multiplets in GCFT/BMSFT can be found in \cite{bootstrap} which focuses on quasi-primaries with non-vanishing boost charge, i.e.  $\xi\neq0$. As we will show later in Section 3, however, our free scalar model also contains a multiplet at the level of primary operators, and it has $\xi=0$.
In the following, we will first review the correlation functions for singlets and multiplets in the $\xi\neq0$ sector, and then provide results for the $\xi=0$ sector.

\subsubsection{Singlets}
The singlet version of BMS primary operators at the origin is defined by \eqref{eigen} and \eqref{highest}.
The operators at other positions can be obtained by acting with the translation operator $U=e^{x L_{-1}+y M_{-1}}$,
\begin{equation}\label{transs}
 O(x,y)=U O(0,0) U^{-1}.
\end{equation}
Using the Baker-Campbell-Hausdorff (BCH) formula, the transformation law for the primary operators are,
\begin{align}
    [L_n,O(x,y)]&=(x^{n+1}\partial_x+(n+1)x^ny\partial_y+(n+1)(x^n\Delta+nx^{n-1}y\xi))O(x,y),\label{Lntrans}\\
    [M_n,O(x,y)]&=(x^{n+1}\partial_y+(n+1)x^n\xi)O(x,y) \quad n \geq -1, \label{Mntrans}
\end{align}
and they can be integrated to derive the transformation laws under the finite transformation \eqref{BMSplane},
\eq{\label{finitetrans}
    \tilde{O}(\tilde{x},\tilde{y})
    &=|f'|^{-\D}\,e^{-\xi\frac{g'+yf''}{f'}}\,O(x,y).
}
By requiring the vacuum to be invariant under the global symmetry, the two-point function (\(G_2\)) and three-point function (\(G_3\)) of primary operators are respectively
\begin{align}
    G_2(x_1,x_2,y_1,y_2) &=d\, \delta_{\Delta_1,\Delta_2}\delta_{\xi_1,\xi_2}|x_{12}|^{-2\Delta_1}e^{-2\xi_1\frac{ y_{12}}{x_{12}}},\\
G_3(x_1,x_2,x_3,y_1,y_2,y_3) &= c_{123}|x_{12}|^{-\Delta_{123}}|x_{23}|^{-\Delta_{231}}|x_{31}|^{-\Delta_{312}}e^{-\xi_{123}\frac{y_{12}}{x_{12}}}e^{-\xi_{312}\frac{y_{31}}{x_{31}}}e^{-\xi_{231}\frac{y_{23}}{x_{23}}}, \label{3ptprimary}
\end{align}
where $d$ is the normalization factor of the two-point function, $c_{123}$ is the coefficient of three-point function which encodes dynamical information of the BMSFTs, and
\begin{equation}
x_{ij}\equiv x_i-x_j,\ \ y_{ij}\equiv y_i-y_j,\ \ \Delta_{ijk}\equiv\Delta_i+\Delta_j-\Delta_k,\ \ \xi_{ijk}\equiv\xi_i+\xi_j-\xi_k.
\end{equation}

\subsubsection{Multiplets}
\if In general, the action of $L_0$ and $M_0$ on the primary operators are not necessarily diagonal. Instead, the representation matrix can be written in the Jordan canonical form. The usual case in consideration is the action of $M_0$ is not diagonal, while that of $L_0$ is still diagonal,
\begin{equation}
[M_0,O]={\bxi}O
\end{equation}
where $O$ are primary operators in the theory, and ${\bxi}$ is block-diagonalized,
\begin{equation}
{\bxi}=
\begin{pmatrix}
\ddots & & & \\
 & \tilde{\xi_i}& & \\
  & & \tilde{\xi_j}& \\
   & & & \ddots\\
\end{pmatrix}
\end{equation}
in which $\tilde{\xi_i}$ are Jordan blocks,
\begin{equation}
\tilde{\xi_i}=
\begin{pmatrix}
 \xi_i& & & \\
  1& \xi_i& & \\
  & \ddots&\ddots &\\
   & & 1& \xi_i\\
\end{pmatrix}_{r\times r}.
\end{equation}\fi

Similar to the discussion of the singlet, local operators corresponding to the highest weight multiplets \eqref{multiplet} are defined by
\begin{equation}
{\bO}(x,y)=U \bO (0,0)U^{-1}, \quad \quad U=e^{x L_{-1}+y M_{-1}}
\end{equation}
 where  $\bO(x,y)$  denotes a multiplet with rank $r$, whose components are denoted by
 $O_{a}(x,y)$ with $a=0,\cdots r-1$.
The BCH formula now leads to the following transformation law,
\begin{align}
[L_n,\bO(x,y)]&=[(x^{n+1}\partial_x+(n+1)x^ny\partial_y)+(n+1)(x^n \Delta+nx^{n-1}y{\bxi})]\bO(x,y),\nonumber\\
[M_n,\bO(x,y)]&=(x^{n+1}\partial_y+(n+1)x^n{\bxi})\bO(x,y),\quad \text{for } n\geq -1, \label{lmmmtrans}
\end{align}
where ${\bxi}$ is the Jordan cell \eqref{jordan}.
Note that the action of $L_0,\,L_{-1}$ and $M_{-1}$ on the highest weight multiplets remains diagonal, while the action of $L_1,\,M_0$ and $M_{1}$ contains non-diagonal parts.
From \eqref{lmmmtrans},  we can get the $T\bO$-OPE and $M\bO$-OPE for the primary multiplet $\bO(x,y)$,
\eq{
T(\tilde{x},\tilde{y}) \bO(x,y) & \sim \frac{\Delta \bO}{(\tilde{x}-x)^2}- \frac{2(\tilde{y}-y) {\bxi} \bO}{(\tilde{x}-x)^3}+\frac{\partial_x \bO}{\tilde{x}-x} -\frac{(\tilde{y}-y)\partial_y \bO}{(\tilde{x}-x)^2} ,\nonumber \\
\label{MOope}M(\tilde{x},\tilde{y}) \bO(x,y) & \sim  \frac{{\bxi} \bO}{(\tilde{x}-x)^2}+\frac{\partial_y \bO}{\tilde{x}-x} .
}
where we organize the expansion in ascending order of the total power of $\left(\tilde{x}-x\right)$ and $\left(\tilde{y}-y\right)$, meanwhile put terms with higher power of $\left(\tilde{y}-y\right)$ behind.

As proved in \cite{Flohr:1997wm,bootstrap},  from an equality for singlets, we can always get the analogue multiplet version by applying the following replacement rule,
\begin{equation}\label{reprule2}
F(\xi,O)\rightarrow \sum_{k=0}^{a}\frac
{1}{k!}\partial_\xi^k F(\xi,O_{a-k})
\end{equation}
where $F(\xi,O)$ denotes any expression which explicitly depends on the operator $O$ and its boost charge $\xi$, and
the replacement should be performed on both sides of the equality.
In particular, the finite transformation law for the multiplet can be obtained from \eqref{finitetrans} by applying this replacement rule, and the result is
\begin{equation}\label{mftran}
    \tilde{O}_{a}(\tilde{x}, \tilde{y})=\sum_{k=0}^{a}\frac{1}{k!}|f'|^{-\D}\,\partial_{\xi}^{k}e^{-\xi\frac{g'+yf''}{f'}}\,O_{a-k}(x,y).
\end{equation}

\paragraph{Quasi-primaries}
Quasi-primary operators transform covariantly under the global part of the BMS$_3$ symmetry, which is isomorphic to the Poincar\'e group in  three dimensions.
The infinitesimal transformations satisfy the rule \eqref{lmmmtrans}, but now with $n=-1,\,0,\,1$, and the OPE with the stress tensor takes the form
\eq{
T(x',y') \bO(x,y) & \sim \cdots+\frac{\Delta \bO}{(x'-x)^2}- \frac{2(y'-y) {\bxi} \bO}{(x'-x)^3}+ \frac{\partial_x \bO}{x'-x}  - \frac{(y'-y)\partial_y \bO}{(x'-x)^2} ,\nonumber \\
M(x',y') \bO(x,y) & \sim \cdots+ \frac{{\bxi} \bO}{(x'-x)^2}+ \frac{\partial_y \bO}{x'-x} .
}
where  $\cdots$ denotes terms more singular than $(x'-x)^{-3}$.  Terms of order $(x'-x)^{-3}$  do not appear due to the conditions coming from $M_1$ and $L_1$.
If $\Delta =2$,
 the scaling of each term in the right hand side must be the same as $(x'-x)^{-4}$ from dimensional analysis. On the other hand, the weight of operators should be bounded from below. This means that the most singular term in the OPE must be of order $(x'-x)^{-4}$.
Further using the relation $L_{-1} M=M_{-1} T$, we conclude that quasi-primary multiplets with $\Delta =2$ have to satisfy
\eq{\label{quasiOPE}
T(x',y') \bO(x,y) & \sim {{\bf c}\over 2(x'-x)^{4}} -{{2\bf c}' (y'-y)\over (x'-x)^{5}} +\frac{\Delta \bO}{(x'-x)^2}- \frac{2(y'-y) {\bxi} \bO}{(x'-x)^3}+ \frac{\partial_x \bO}{x'-x}  - \frac{(y'-y)\partial_y \bO}{(x'-x)^2} ,\nonumber \\
M(x') \bO(x,y) & \sim {{\bf c}' \over 2(x'-x)^{4}}  + \frac{{\bxi} \bO}{(x'-x)^2}+ \frac{\partial_y \bO}{x'-x},
}
  where ${\bf c}$ and $ {\bf c}' $ are constant vectors with the same rank as $\bO$.
From the OPE \eqref{BMSope}, it is straightforward to see that the stress tensor $(2M, \, T )$ form a rank-two multiplet, with conformal weight $\Delta=2$, and boost charge $\bxi=\left(\begin{matrix}
   0&0  \\
   1&0\end{matrix}\right)$.
In this case, we have \eq{{\bf c}=(c_M,\, c_L)^T,\quad {\bf c}' =(0,\,2c_M)^T.}

\paragraph{Two-point functions}

Let us first consider two-point functions $\langle O_{ia} O_{jb}\rangle$, where $O_{ia}$ belongs to a rank-$r_i$ multiplet $\bO_{i}$, and $O_{jb}$ belongs to a rank-$r_j$ quasi-primary multiplet $\bO_{j}$.
The two point functions can be determined by the Ward identities with respect to global symmetries. It is always possible to choose a basis so that the   operators belonging to different multiplets have vanishing two-point functions.
The Ward identities with $M_{-1}$ and $L_{-1}$ imply the two point functions only depends on $x_{12}$ and $y_{12}$. Further using Ward identities with $M_0, ~M_{1},~L_0$ and $ L_{1}$, we get
\begin{equation}
\langle O_{ia}(x_1,y_1)O_{jb}(x_2,y_2)\rangle=\delta_{ij}\,|x_{12}|^{-2\Delta_i}e^{-2\xi_i \frac{y_{12}}{x_{12}}} D_{ab}({y_{12}\over x_{12}}),\label{ansatz}
\end{equation}
where the function $D_{ab}({y_{12} \over x_{12}})$
satisfies the following differential equations
\eq{
& D'_{ab}+D_{a-1,b}+D_{a,b-1}=0,  \label{m0diff}\\
&(x_{1}+x_2)D'_{ab}+2 x_1 D_{a-1,b}+2 x_2 D_{a,b-1}=0.
} In the above equation we have omitted the argument of  $D_{ab}$, and prime denotes derivative with respect to the argument ${y_{12} \over x_{12}}$.
Combing the above two equations, we learn that $D_{ab}$ depends on the label $a$ and $b$ only through the the sum $a+b$,
\eq{D_{a,b-1}=D_{a-1,b}=\cdots =D_{b,a-1}=D_{b-1,a}, \quad a,\,b=1,\cdots r-1\label{sym}}
and $D_{ab}$ vanishes whenever one of the indices is $0$,
\eq{D_{a0}=D_{0b}=0, \quad a,\,b=0\cdots r-2\label{initial}}
where $r_i=r_j=r$.
Denote $q\equiv a+b+1-r$, and then the most general solution of \eqref{m0diff} satisfying the conditions \eqref{sym} and \eqref{initial} is then given by
\begin{equation}\label{sdab}
D_{ab}=
\left\{\begin{array}{ll}
0& \quad\mbox{for $q\equiv a+b+1-r<0$}\\
\, \sum\limits_{k=0}^{q}d_{(r+q-k)}\,\frac{1}{k!}\,(\frac{-2y_{12}}{x_{12}})^k& \quad \mbox{for $q\ge0$},\end{array} \right.
\end{equation}
where $d_{(r+q-k)}$ are $r$ undetermined integration constants, which can be further fixed by redefining the operators in the multiplet.
To do so, we need to find the most general linear transformations $\bO\rightarrow R\bO$, where $R$ is a $r\times r$ matrix, that leave the Jordan cell $\bxi$ invariant, namely \begin{equation}
{\bxi}R=R{\bxi}
\end{equation}
The solution can be written as
\begin{equation}
R_{ab}=
\begin{cases}
c_{a-b}&\mbox{if $a\geq b$}\\
0&\mbox{otherwise},
\end{cases}
\end{equation}
which contains $r$ arbitrary constants $c_k,\ k=0,\cdots,r-1$.
These $(r-1)$ independent parameters $c_k,\,k=1,\cdots, r-1$ can be used to eliminate $r-1$ degrees of freedom in \eqref{sdab}, and leave an overall normalization related to the diagonal element $c_0$.
This simplifies the two-point functions to the following canonical form,
\begin{equation}\label{2pt}
\langle O_{ia}(x_1,y_1)O_{jb}(x_2,y_2)\rangle =\left\{\begin{array}{ll}
0& \mbox{for $q<0$}\\
\, \delta_{ij}\,  d_r\,  |x_{12}|^{-2\Delta_i} e^{-2\xi_i\frac{y_{12}}{x_{12}}}\frac{1}{q!}\left(-\frac{2y_{12}}{x_{12}}\right)^q,& \mbox{otherwise},\end{array} \right.
\end{equation}
where $d_r$ is the overall normalization.

\paragraph{In-states and out-states}

Using the state-operator correspondence\footnote{See appendix A for more details of radial quantization and the state-operator correspondence.} on the plane,
 \begin{equation}\label{operator-state}
|O_{a}\rangle=\lim\limits_{ y\rightarrow 0 \atop x\rightarrow 0}O_{a}(x,y)|0\rangle
\end{equation}
we can define out-states as the Hermitian conjugate of the operator inserted at infinity,
\begin{equation}\label{outstate}
\langle O_{a}|=\lim\limits_{ y\rightarrow 0 \atop x\rightarrow \infty}\sum\limits_{k=0}^{a}\langle 0|O_{a-k}(x,y) \frac{1}{k!} \partial_{\xi}^k e^{2\xi\frac{y}{x}} x^{2\Delta}
\end{equation}
where we have used the transformation rule \eqref{mftran} to move the operator from the origin to the point $x=\infty, \, y=0$. Note that the operators on the right hand side of \eqref{outstate} are to be understood as acting to the right.
For all singlets including the vacuum, the definition of the out-state \eqref{outstate}  is the same as in CFT$_2$. For multiplets, however, the out-state $\langle O_{a}|$ becomes a mixture of operators in the multiplet with indices no bigger than $a$.

The inner product between different components of a rank-$r$ multiplet primary can then be calculated using \eqref{2pt} as,
\eq{
\langle O_a|O_b\rangle &=\lim\limits_{x_1\to \infty,x_2 \to 0, \atop  y_{1}\to 0, y_{2}\to 0}\frac{1}{k!} \partial_{\xi}^k e^{2\xi\frac{y_1}{x_1}} x_1^{2\Delta} \sum_{k=0}^{a}\langle O_{a-k}(x_1,y_1) O_{b}(x_2,y_2)\rangle \nonumber\\
&= \delta_{a+b,r-1} \label{norm}
}
Unlike the case for singlets,  for multiplets the two-point function \eqref{2pt} is different from the inner product \eqref{norm}. This is a characteristic feature of multiplets.
From \eqref{norm}, we can see that the $r$-dimensional  matrix $\langle O_a|O_b\rangle$ has two different eigenvalues $\pm 1$,  within which the eigenvalue $-1$ has algebraic multiplicity $\lfloor r/2 \rfloor$.
This means that if the theory contains highest weight multiplets, there must be primary states with negative norm in the theory.
Note that this is to be distinguished from earlier discussions of unitarity for the highest weight singlets \cite{2009}, where descendent states with negative norms have been found assuming primary states have positive norms. Thus, we have found another indication that BMSFT in the highest weight representation is not unitary. In section 6, we will introduce a dual basis which is linear combination of the out-state \eqref{outstate} so that their inner products with the in-states are diagonal. The dual basis is useful to define the trace.

\paragraph{Three-point functions}
The three-point functions involving multiplets can also be determined by the Ward identities.
For three multiplets $\bO_i,\,i=1,2,3$, the general form of the three point function is given by,
\begin{equation}
\langle O_{ia}O_{jb}O_{kc}\rangle =A \,B\, C_{ijk}\label{3pf}
\end{equation}
where $i,\,j$ and $k$ label the multiplets, while $a,\,b$ and $c$ label the components within a multiplet, and
\begin{eqnarray}
A&=&\exp(-\xi_{123}\frac{y_{12}}{x_{12}}-\xi_{312}\frac{y_{31}}{x_{31}}-\xi_{231}\frac{y_{23}}{x_{23}}),\\
B&=&|x_{12}|^{-\Delta_{123}}|x_{23}|^{-\Delta_{231}}|x_{31}|^{-\Delta_{312}},\\
C_{ijk;abc}&=&\sum_{n_1=0}^{a-1}\sum_{n_2=0}^{b-1}\sum_{n_3=0}^{c-1}c_{ijk}^{(n_1 n_2n_3)}\frac{(p_i)^{n_1}(p_j)^{n_2}(p_k)^{n_3}}{n_1!n_2!n_3!},
\end{eqnarray}
with
\begin{equation}
p_i=\partial_{\xi_i}\ln A.
\end{equation}
Note that the three point function \eqref{3pf} factorizes into a term $A$ which depends on the boost charges, a term $B$ which depends on the conformal weights,  and a structure constant term $C_{ijk;abc}$. Both $A$ and $B$ are determined by kinematics, while $C_{ijk;abc}$ encodes interactions.
Note that $O_{ia},O_{jb},O_{kc}$ can belong to different multiplets of rank $r_1,r_2,r_3$ respectively. The coefficient $c_{ijk}^{(n_1n_2n_3)}$ encodes the dynamical information of the theory. When $r_1=r_2=r_3=1$, the three-point function reduces to \eqref{3ptprimary}.

\subsection*{Multiplets with $\xi=0$}
The case with $\xi=0$ turns out to be a bit subtle. On the one hand, the action of $M_0\sim \partial_y$ acts trivially on singlets with $\xi=0$.
As a result, correlation functions for singlets with $\xi=0$ do not depend on $y$ and they reduce to correlators in chiral CFTs.

On the other hand, if multiplets exist, the action of $M_0$ is still non-trivial because of existence of off-diagonal elements, and we expect the correlators to  have non-trivial dependence on $y$. To calculate
the correlation functions  at $\xi=0$, we  should take the $\xi\rightarrow0$ limit of the correlation functions of the $\xi\neq0$ case. If there are derivatives with respect to $\xi$, such as in the three point function \eqref{3pf}, one should take the derivatives first and then take $\xi\rightarrow0$.

As an example,  the two-point correlation functions for rank-2 multiplet with $\xi=0$ can be written as,
\begin{eqnarray}\label{xi02pt}
\nonumber\langle O_1O_1 \rangle&=&-\frac{1}{x^{2\Delta}}\frac{2y}{x} \\
\nonumber\langle O_0O_1 \rangle&=&\frac{1}{x^{2\Delta}} \\
\langle O_0O_0 \rangle&=&0
\end{eqnarray}
In section 3, we will see that such a multiplet with $\xi=0$ appears in our free scalar model.

To recapitulate, in this section we have introduced the multiplet version of the highest weight representation of the BMS algebra. The action of $M_0$ on the primary states  is block diagonal, as described by \eqref{multiplet}.
The OPE between a rank $r$ multiplet and the stress tensor \eqref{MOope} also acquires a non-diagonal term, which mixes different components within the multiplet. Further more, two-point \eqref{2pt} and three-point \eqref{3pf} correlation functions  for multiplets have also been worked out using the BMS symmetry. It is very interesting to further study four point functions, which allows a formulation of bootstrap program with BMS multiplets, similar to the discussion of BMS bootstrap for singlets \cite{Bagchi:2017cpu}. The results for multiplets are more complicated, and related discussions will be reported elsewhere \cite{BinChen03}.

\section{BMS Free Scalar Model}
In this section we study a free scalar model which has BMS invariance.
In section 3.1 we introduce the classical theory and discuss the classical BMS invariance.
In section 3.2 we perform canonical quantization and in section 3.3 we discuss primary operators and correlation functions.

\subsection{The classical theory}
We start with the action on a cylinder parameterized by$(\sigma,\tau)$ with $\sigma\sim \sigma+2\pi$,
\eq{
S = \frac{1}{4\pi} \int d\sigma d\tau \left( \p_{\tau} \phi \right)^2. \label{action}
}
The action also appears as part of the worldsheet action in a tensionless limit of string theory \cite{Schild, Isberg:1993av, Bagchi, Casali:2016atr}.
In this paper, however, we will study the model \eqref{action} as a quantum field theory itself, without embedding it into a larger theory.
The equation of motion reads
\begin{equation}
\partial_{\tau}^2 \phi=0.	\label{eom}
\end{equation}
The solutions to the equation of motion \eqref{eom} satisfying the periodic boundary condition on the cylinder can be written in terms of the mode expansion as
\begin{equation}
\phi(\sigma,\tau)=\sum_{n=-\infty}^\infty e^{-i \sigma n}(A_n+i\tau  B_n ).	\label{modecylinder}
\end{equation}
The reality condition then implies the adjoint relation
\begin{equation}\label{m34c}
A^\dagger_n=A_{-n},\ \ \ B^\dagger_n=-B_{-n}.
\end{equation}
The conjugate momentum to the field $\phi$ is given by
 \eq{\Pi ={\delta S\over \delta {\partial_\tau \phi}}= \frac{1}{2\pi}\partial_{\tau}\phi,}
and the Poisson bracket is
\eq{\left\{ \phi\left(\sigma,\tau \right), \Pi\left(\sigma^{\prime},\tau \right)\right\} = \delta(\sigma-\sigma^{\prime}).\label{Possion}}

It is not difficult to check that this action is invariant under the BMS transformations \eqref{BMSfinite}. For infinitesimal transformations parametrized by $\tilde{\varepsilon}(\sigma)$ and $\varepsilon(\sigma)$,
\begin{align}
\sigma\to \sigma^{\prime} &= \sigma + \varepsilon(\sigma),\\
\tau \to \tau^{\prime} &= \tau + \varepsilon^{\prime}(\sigma)\tau + \tilde{\varepsilon}(\sigma),   \label{infinitesimal_bms}
\end{align}
the field $\phi$ transforms as a scalar,
\eq{\label{transPhi}
\delta_{\varepsilon(\sigma)} \phi &= -\varepsilon(\sigma) \p_\sigma \phi -\varepsilon'(\sigma) \tau \p_\tau \phi,\nonumber\\
\delta_{\tilde{\varepsilon}(\sigma)} \phi &= -\tilde{\varepsilon}(\sigma) \p_\tau \phi.
}
The corresponding Noether currents can be obtained from the standard Noether procedure,
\if
The conservation laws associated with $\varepsilon(\sigma)$ and $\tilde{\varepsilon}(\sigma)$ symmetries are just the vanishing of the square brackets, we repeat them here for later convenience, The Noether currents can be compactly written as 1-forms,
\eq{
\p_\tau\big( 2 \varepsilon(\sigma) \p_\sigma \phi \p_\tau \phi + \varepsilon'(\sigma) \tau \p_\tau\phi \p_\tau\phi \big) + \p_\sigma \big( -\varepsilon(\sigma) \p_\tau\phi \p_\tau\phi \big) &\overset{\text{on-shell}}{=} 0,\\
\p_\tau \big(\tilde{\varepsilon}(\sigma) \p_\tau\phi \p_\tau\phi \big) &\overset{\text{on-shell}}{=} 0.
}
The conserved current with $\varepsilon(\sigma)$ and $\tilde{\varepsilon}(\sigma)$ can also be read out,
\eq{
4\pi \pmb{j}_{\varepsilon(\sigma)} &= \big( 2 \varepsilon(\sigma) \p_\sigma \phi \p_\tau \phi + \varepsilon'(\sigma) \tau \p_\tau\phi \p_\tau\phi \big) d\sigma  + \big( \varepsilon(\sigma) \p_\tau\phi \p_\tau\phi \big) d\tau,\\
4\pi \pmb{j}_{\tilde{\varepsilon}(\sigma)} &= \big( \tilde{\varepsilon}(\sigma) \p_\tau\phi \p_\tau\phi \big) d\sigma.
}
\fi

\eq{
2\pi \pmb{j}_{\varepsilon(\sigma)} &= -\big( \varepsilon(\sigma) T + \varepsilon'(\sigma) \tau M \big) d\sigma  - \varepsilon(\sigma) M d\tau,\\
2\pi \pmb{j}_{\tilde{\varepsilon}(\sigma)} &= -\big( \tilde{\varepsilon}(\sigma) M \big) d\sigma,
}
where the currents  \(T\) and \(M\)  are defined as
\begin{align}
T &= -\partial_{\sigma}\phi\partial_{\tau}\phi,\\
M &= -\frac{1}{2}\partial_{\tau}\phi\partial_{\tau}\phi.
\end{align}
The conservation laws are given by
\eq{
d\pmb{j}_{\varepsilon(\sigma)} =d\pmb{j}_{\tilde{\varepsilon}(\sigma)} =0,
}
which can also be expressed in terms of the currents as
\eq{
\p_\tau T=\p_\sigma M,\ \ \partial_\tau M=0. \label{conservation}
}
The conservation laws allow us to define the conserved charges as
\eq{\label{classicalcharges}
Q_{\varepsilon(\sigma)} &=  \int_{\sigma\text{-cycle}} \pmb{j}_{\varepsilon(\sigma)} =-\frac{1}{2\pi} \int_0^{2\pi} d\sigma \big( \varepsilon(\sigma) T + \varepsilon'(\sigma) \tau M \big) \sim -\frac{1}{2\pi}\int_0^{2\pi} d\sigma ~\varepsilon(\sigma) \big( T - \tau \p_\sigma M \big),\nonumber\\
Q_{\tilde{\varepsilon}(\sigma)} &= \int_{\sigma\text{-cycle}} \pmb{j}_{\tilde{\varepsilon}(\sigma)} = -\frac{1}{2\pi} \int_0^{2\pi} d\sigma  ~\tilde{\varepsilon}(\sigma) M .
}

In additional to the BMS symmetries, we also note that there is an affine $U(1)$ symmetry, realized by $\tau$-independent shifts of the field $\phi$ parametrized by $\Lambda(\sigma)$,
\eq{ \label{internal_U1}
\phi(\sigma,\tau) \to \tilde{\phi}(\sigma,\tau)=\phi(\sigma,\tau) + \Lambda(\sigma).
}
The associated Noether current and conserved charge are
\eq{\label{jcurrent}
2\pi \pmb{j}_{\Lambda(\sigma)} &= i\Lambda(\sigma) J(\tau,\sigma) d\sigma,\\
J(\tau,\sigma)&=J(\sigma)= i\p_\tau \phi	\label{Jcurrent},\\
Q_{\Lambda(\sigma)} &= \int_{\sigma\text{-cycle}} \pmb{j}_{\Lambda(\sigma)} = \frac{i}{2\pi} \int_0^{2\pi} d\sigma \Lambda(\sigma) J(\sigma).\label{jcharge}
}
Interestingly, we note that the current $J$ is proportional to the canonical momentum $\Pi$, and that its Sugawara stress tensor is proportional to the current $M$,
\eq{\Pi=-{i\over2\pi} J, ~~ M={1\over 2}J^2.}

As a consistency check, we find these charges indeed implement the transformation \eqref{transPhi} and \eqref{internal_U1} via the Poisson bracket \eqref{Possion},
\eq{
\{Q_{\varepsilon(\sigma)},\, \phi(\tau,\sigma) \}&=-\varepsilon(\sigma) \partial_{\sigma}\phi-\tau \varepsilon'(\sigma) \partial_{\tau}\phi  =\delta_{\varepsilon(\sigma)} \phi(\tau,\sigma),\\
\{Q_{\tilde{\varepsilon}(\sigma)},\, \phi(\tau,\sigma) \}&=-\tilde{\varepsilon}(\sigma) \p_\tau \phi=\delta_{\tilde{\varepsilon}(\sigma)} \phi(\tau,\sigma),\\
\{Q_{\Lambda(\sigma)},\, \phi(\tau,\sigma)\}&=\Lambda(\sigma).\label{u1charge}
}
Furthermore, the currents transform as
\begin{align}
\delta_{\varepsilon,\,\tilde{\varepsilon},\,\Lambda} T (\tau,\sigma) =&-2 \varepsilon' T-\varepsilon T'-\varepsilon' \tau \p_\tau T-2\varepsilon''\tau M - 2\tilde{\varepsilon}^{\prime}M  -\tilde{\varepsilon}M^{\prime} +i\Lambda J',\\
\delta_{\varepsilon,\,\tilde{\varepsilon},\,\Lambda} M (\tau,\sigma)  =&-2\varepsilon' M- \varepsilon M',\\
\delta_{\varepsilon,\,\tilde{\varepsilon},\,\Lambda} J (\tau,\sigma)  =& -\varepsilon' J-\varepsilon J'.
\end{align}
To find the symmetry algebra, we need to
first find the symmetry generators, which are mode expansion of the conserved charges \eqref{classicalcharges} and \eqref{jcharge},
and
can be obtained by
expanding the symmetry parameters $\varepsilon(\sigma), \tilde{\varepsilon}(\sigma)$ and $\Lambda(\sigma)$ in terms of the Fourier modes,
\eq{
\varepsilon_n = \tilde{\varepsilon}_n =e^{in\sigma} , \quad \Lambda_n = ie^{i n \sigma}.
}
The resulting symmetry generators are
\eq{\label{cylindergen}
L_{n} &\coloneqq  Q_{\varepsilon_n} = -\frac{1}{2\pi}\int_0^{2\pi} d\sigma \varepsilon_n \big( T - \tau \p_\sigma M \big),\nonumber\\
M_{n} &\coloneqq Q_{\tilde{\varepsilon}_n} =  -\frac{1}{2\pi} \int_0^{2\pi} d\sigma  \tilde{\varepsilon}_n M ,\\
J_{n} &\coloneqq Q_{\Lambda_n} = -\frac{1}{2\pi} \int_0^{2\pi} d\sigma \Lambda_n \p_\tau \phi.\nonumber
}
These charges form an algebra under Poisson bracket,
\begin{align}&\{Q_{\varepsilon_m(\sigma)},\, Q_{\varepsilon_n(\sigma)}\}=-i\left(m-n\right)Q_{\varepsilon_{m+n}(\sigma)} \\
&\{Q_{\varepsilon_m(\sigma)},\, Q_{\tilde\varepsilon_n(\sigma)} \} =-i\left(m-n\right)Q_{{\tilde\varepsilon}_{m+n}(\sigma)}  \nonumber\\
&\{Q_{\tilde\varepsilon_m(\sigma)} ,\, Q_{\tilde\varepsilon_n(\sigma)}  \} =0\nonumber\\
&\{Q_{\varepsilon_m(\sigma)} ,\, Q_{\Lambda_n(\sigma)}  \} = inQ_{\Lambda_{m+n}(\sigma)} \nonumber\\
&\{Q_{\tilde\varepsilon_m(\sigma)},\, Q_{\Lambda_n(\sigma)}  \}=0 ,\\
& \{Q_{\Lambda_m(\sigma)} ,\,Q_{\Lambda_n(\sigma)} \}= 0\nonumber \label{classical_alg}
\end{align}
Under the canonical quantization replacement $\{\cdot\,,\cdot\}\to -i[\cdot\,,\cdot],$
we get the BMS algebra without central terms at the classical level, namely
\begin{align}
& [L_m,\, L_n ]  =\left(m-n\right)L_{m+n},\nonumber \\
& [L_m,\, M_n ] =\left(m-n\right)M_{m+n},  \\
& [M_m,\, M_n ] =0.\nonumber
\end{align}
Note that the affine $U(1)$ symmetry and the Virasoro algebra generated by $L_n$'s together form a Virasoro-Kac-Moody algebra,
\begin{eqnarray}
  && [L_m,\, J_n ]  =-nJ_{m+n}, \nonumber\\
  && [M_m, \,J_n]=0, \label{vir-kac}\\
  &&  [J_m, \,J_n]=0.\nonumber
\end{eqnarray}
\subsection*{Mapping to the plane}
Under the plane to cylinder map \eqref{plane2cylinder}, the solution \eqref{modecylinder} to the equation of motion rewritten on the plane is
\begin{equation}\label{mode}
\phi(x,y)=A(x)+yB(x),
\end{equation}
where
\begin{equation}
A(x)=\sum\limits_{n} A_n x^{-n},\ \ \ B(x)=\sum\limits_{n} B_n x^{-n-1}.
\end{equation}
The Noether currents  \(T\) and \(M\) corresponding to translations along \(x\) and \(y\) are
\begin{align}
T &= -\partial_{x}\phi\partial_{y}\phi,\\
M &= -\frac{1}{2}\partial_{y}\phi\partial_{y}\phi.
\end{align}
The conserved charges on the plane are
\eq{\label{fchar}
L_{n} &= \frac{1}{2\pi i}\oint  \left((n+1)x^{n} y M  + x^{n+1} T\right) dx, \nonumber\\
    M_n&= \frac{1}{2\pi i }\oint dx x^{n+1} M.
}
The internal $U(1)$ current is
\eq{
    J= \partial_y\phi.} \label{cylinchar}
    and the charges are given by
\begin{align}\label{Jplane}
J_n=\frac{1}{2\pi i} \int dx x^{n} \partial_y\phi.
\end{align}
As a consistency check, theses charges on the plane \eqref{fchar} and \eqref{Jplane} are consistent with those on the cylinder \eqref{cylindergen}  under the coordinate transformation \eqref{plane2cylinder}.

\subsection{Canonical Quantization}
In this subsection we perform canonical quantization to the scalar model \eqref{action}.
This amounts to replacing the Poisson bracket \eqref{Possion} with the canonical commutation relation
\begin{equation}
[\phi (\sigma_1,\tau_0),\,\Pi(\sigma_2,\tau_0)]=i\delta(\sigma_1-\sigma_2)=i\sum_n \frac{1}{2\pi}e^{-in(\sigma_1-\sigma_2)}.
\end{equation}
which can be equivalently written in terms of the mode operators
\begin{equation}\label{c347r}
[A_n,B_m]=\delta_{n+m,0},\quad
[A_n,A_m]=[B_n,B_m]=0.
\end{equation}
The commutation relations \eqref{c347r} are valid both on the cylinder and on the plane.
Henceforth later discussions will be carried out on the plane, unless otherwise specified.

The quantum version of the classical Noether currents $T(x,y)$,\,$M(x)$ and $J(x)$ that generate translations along $x$ and $y$ and the internal $U(1)$ symmetry now become operators,
\begin{equation}
T(x,y)=-:\partial_x\phi\partial_y\phi:,\ \  M(x)=-\frac{1}{2}:\partial_y\phi\partial_y\phi:,\ \ J(x)=\partial_y\phi,
\end{equation}
where the definition of the normal order $:\cdots:$ depends on the choice of the vacuum, which will be specified momentarily. Here we would like to keep the normal ordering implicit.
The currents can be expanded in Laurent series as
\eq{
T&=\sum_{n} L_n x^{-n-2}-\sum_n (n+1)yM_{n-1}x^{-n-2},	\label{T_modes}\\
M&=\sum_n M_n x^{-n-2},	\label{M_modes}\\
J&=\sum_n J_n x^{-n-1}.
}
which can be inverted to define infinitely many charges as
\eq{
L_n&=\sum_{k=-\infty}^\infty :kA_kB_{n-k}:,\quad M_n=-\frac{1}{2}\sum_{k=-\infty}^\infty :B_kB_{n-k}:,\ \ \ J_n=B_n,	\label{pl_in_char}
}
with the Hermitian conjugates given by
\eq{L_n^\dagger=L_{-n},\quad M_n^\dagger=M_{-n},\quad J_n^\dagger=-J_{-n}.}
The charges are the quantum version of the classical charges \eqref{fchar} and \eqref{Jplane} on the plane.

\subsubsection{Vacuum in the highest weight representation and the state space}
Note that so far we have not specified the vacuum for the model \eqref{action}.
We are interested in the vacuum this is invariant under the global symmetries generated by $L_{0,\pm1}$ and $M_{0,\pm1}$. That is, the vacuum has to satisfy
\eq{
L_{\pm 1, \,0}|0\rangle=M_{\pm 1, \,0}|0\rangle=0\label{globalann}
}
To describe the vacuum in canonical quantization, we need to translate these conditions in terms of $A_n$'s and $B_m$'s.
As $[A_n, B_{-n}]=1$, $A_n$ and $B_{-n}$ cannot annihilate the vacuum simultaneously.
If we let $B_{k_0}|0\rangle=0$ for a given positive integer $k_0$, then $A_{-k_0}|0\rangle\neq0.$
Note that the expressions of $L_{\pm 1}$ \eqref{pl_in_char} contain a term $-k_0 A_{-k_0} B_{k_0 \pm 1}$. If we require $L_{\pm 1}$ to annihilate the vacuum term by term, $B_{k_0 \pm 1}$ has to annihilate the vacuum. We can keep using this argument
until we arrive at $B_0$. Similar arguments also apply to other cases, and we learn that
\begin{itemize}
\item[I.] if there exists some positive $ k_0$ s.t. $B_{k_0}|0\rangle=0$ , then $B_{k}|0\rangle=0$ for all $k\ge0$;
\item[II.] if there exists some positive $ k_0$ s.t. $B_{-k_0}|0\rangle=0$, then $B_{-k}|0\rangle=0$ for all $k\ge0$;
\item[III.] if there exists some positive $ k_0$ s.t. $A_{k_0}|0\rangle=0$, then $A_{k}|0\rangle=0$ for all $k>0$;
\item[IV.] if there exists some positive $ k_0$ s.t. $A_{-k_0}|0\rangle=0$, then $A_{-k}|0\rangle=0$ for all $k>0$.
\end{itemize}
Obviously, neither $I$ and $IV$  nor $II$ and $III$ can happen simultaneously.
Therefore there are altogether two physically different choices \footnote{The other two choices can be obtained from these two by switching the sign. } for the vacuum that is compatible with both the commutation relations and the symmetry condition \eqref{globalann}.
Let us start with entry I listed above, namely
\eq{B_k|0\rangle=0,\quad A_{-k}|0\rangle\neq0,\quad \forall k\geq 0.\label{choiceI}}
Then we have two feasible choices, case II or case III. Choosing I and II leads to the so-called induced vacuum, which we will describe in detail in appendix B, while choosing I and III leads to the highest weight vacuum, as discussed below.

Here in this section, we will focus on the choice with I and III, which combine as the following conditions,
\eq{&\label{vac1.1}
A_n|0\rangle=0,\ n>0,\\
 &
B_n|0\rangle=0,\ n\geq0\nonumber.
}
It leads to the following normal ordering prescription via creation and annihilation operators,
\begin{eqnarray}\label{normal}
:A_nB_m:=
\begin{cases}
  A_nB_m, &  n\leq0 \\
  B_mA_n, & n>0.
\end{cases}
\end{eqnarray}
From the vacuum condition \eqref{vac1.1}, and the normal ordering \eqref{normal}, it is not difficult to verify that on the plane
\begin{equation}\label{JM0}
M_n|0\rangle=0,  \quad L_n|0\rangle=0, n \geq -1.
\end{equation}
In other words, the vacuum \eqref{vac1.1} is
i) a highest weight state (singlet) with zero conformal weight and boost charge and ii) invariant under the global part of the BMS algebra.
Therefore the choice of the vacuum \eqref{vac1.1} is the proper vacuum in the highest weight representation.

Note that $J_n=B_n$, then \eqref{vac1.1} and  \eqref{JM0} imply that this vacuum is also the highest weight vacuum of the Virasoro-Kac-Moody algebra \eqref{vir-kac}.
If the free scalar BMSFT is part of the worldsheet theory of tensionless strings, the vacuum \eqref{vac1.1} is the vacuum of a single string with zero momentum in the $\phi$ direction of the target space.
The momentum can be turned on by considering an eigenstate of $B_0$ which satisfies
\eq{
&J_0 |\alpha\rangle=B_0 |\alpha\rangle =\alpha |\alpha\rangle,\\
&A_n|\alpha\rangle=0,\quad B_n|\alpha\rangle=0,\quad n>0\nonumber.
}
This can be viewed as a coherent state in the Fock space basis, and one can check that this state is also a highest weight state with
\eq{
\Delta=0,\quad \xi=-{\alpha^2\over2}.}
For $\alpha\neq0,$ the momentum eigenstate are not annihilated by translational generators, $M_{-1}|\alpha\rangle\neq0,  \,L_{-1}|\alpha\rangle\neq0$.
Other states in the theory can be obtained by acting creation operators on zero mode states $|\alpha\rangle $.

Putting everything together, we now describe the state space of the free BMS scalar with the choice of the vacuum \eqref{vac1.1}.
Let $\vec{i}\equiv (i_1,\,i_2\,\cdots),\, \vec{i}\equiv (j_1,\,j_2\,\cdots)$, then the state space is spanned by
\begin{equation}\label{basis}
|\vec{i},\vec{j}; \alpha\rangle :=A_{-1}^{i_1}A_{-2}^{i_2}\cdots B_{-1}^{j_1}B_{-2}^{j_2}\cdots|\alpha\rangle.
\end{equation}

\subsubsection{The quantum BMS algebra }
Now we consider the action of other charges on the vacuum, and calculate the resulting algebra.
With some straightforward but tedious calculation,
we find that the  generators \eqref{pl_in_char}  indeed form the BMS algebra \eqref{BMSalg} with central charges
\eq{c_L=2, \quad c_M=0.}
Additionally, the U(1) charges $J_n$ together with $L_n$ form a $U(1)$ Virasoro-Kac-Moody algebra \eqref{vir-kac}, with vanishing Kac-Moody level.

Before moving on, we briefly comment on operators on the plane versus on the cylinder.
On the plane,  the normal ordering \eqref{normal} implies the vacuum expectation values of the stress tensors on the plane vanishes
\begin{equation}
\langle T(x,y)\rangle=\langle M(x,y)\rangle=0
\end{equation}
and hence all the vacuum expectation values of the BMS charges are zero.
Using the transformation law \eqref{translaw} under the plane-to-cylinder map \eqref{plane2cylinder},
 the zero-mode generator of the Virasoro algebra on the cylinder has a shift,
\begin{equation}
L_0^{cyl}=L_0^{pl} -\frac{1}{12}.\label{cylindershift}
\end{equation}
The above results can also be obtained by assuming symmetric ordering in $L_0^{cyl}$, then the Casimir energy can then be obtained  from $\zeta$-function regularization.

\subsection{Primary operators and Correlation functions}

In this subsection we calculate the Green's function, list the fundamental primary operators, and calculate their correlation functions.

With the mode expansion \eqref{mode} of the fundamental field $\phi$ and the choice of the vacuum in the highest weight representation \eqref{vac1.1}, we can define the Green's function of $\phi$ as,
\begin{equation}\label{2ptdef}
\langle\phi(x_1,y_1)\phi(x_2,y_2)\rangle=\langle 0|X(\phi(x_1,y_1)\phi(x_2,y_2))|0\rangle-\langle 0|:\phi(x_1,y_1)\phi(x_2,y_2):|0\rangle
\end{equation}
where $X(\cdots)$ denotes radial order on the complexified $x$-plane, related to the time order on the Lorentz cylinder, as explained in the appendix A. Additionally, $:\cdots:$ denotes the normal order \eqref{normal} which is compatible with the highest weight vacuum \eqref{vac1.1}.
From the commutation relation of the modes $A_n$ and $B_n$, we learn that only the cross terms contribute, and that the Green's function is given by
 \begin{eqnarray}
\langle\phi(x_1,y_1)\phi(x_2,y_2)\rangle   = -\frac{y_1-y_2}{x_1-x_2}.	\label{green}
\end{eqnarray}

This provides the following OPE for the fundamental field
\eq{
\phi(x_1,y_1)\phi(x_2,y_2)\sim-\frac{y_1-y_2}{x_1-x_2}\label{phiphi}
}
The OPEs of other operators can then be obtained from \eqref{phiphi} via Wick contractions. In particular, we note that the OPEs among the stress tensors read,
\eq{\label{TMope}
T(x',y') T(x,y) & \sim \frac{1}{(x'-x)^4}+ \frac{2T(x,y)}{(x'-x)^2}- \frac{4(y'-y) M(x,y)}{(x'-x)^3}  +\frac{\partial_x T(x,y)}{x'-x}-\frac{(y'-y)\partial_y T(x,y)}{(x'-x)^2} ,\nonumber\\
T(x',y') M(x,y) & \sim \frac{2M(x,y)}{(x'-x)^2}+\frac{\partial_x M(x,y)}{x'-x},\nonumber\\
M(x',y') M(x,y) & \sim 0.
}
These OPEs are consistent with the BMS algebra \eqref{BMSalg}, which has been calculated directly in the previous subsection from the mode expansion of $T$ and $M$ in terms of $L_n$ and $M_n$ and the commutation relation of $A_n$ and $B_n$. In particular, the central charges $c_L=2$ and $c_M=0$ can be read from the most singular terms.

\subsubsection{Primary operators}
Now let us find the primary operators in the free BMS scalar field theory. According to the general discussion in section 2, BMS Primary operators in a generic multiplet have the defining property that the OPEs with the stress tensors $T$ and $M$ have to be of the form \eqref{MOope}. This can be used to find all the primary operators in the free BMS scalar field theory.
We first consider
\begin{equation}
O_0(x,y)\equiv i\partial_y \phi(x,y), \,O_1(x,y)\equiv i\partial_x \phi(x,y).\label{dphi}
\end{equation}
where the pre-factor $i$ makes the operators Hermitian. Their OPEs with the stress tensor read
\eq{
T(x',y')O_0(x,y)&=\frac{O_0}{(x'-x)^2}+\frac{\partial_x O_0}{x'-x},\\
T(x',y')O_1(x,y)&=\frac{O_1}{(x'-x)^2} +\frac{\partial_x O_1}{x'-x} +  \frac{-2(y'-y)O_0}{(x'-x)^3}+ \frac{-(y'-y)\partial_y O_1}{(x'-x)^2}, \nonumber\\
M(x',y')O_0(x,y)&=0,\nonumber\\
M(x',y')O_1(x,y)&=\frac{O_0}{(x'-x)^2}+\frac{\partial_y O_1}{x'-x}. \nonumber
}
Comparing to \eqref{MOope}, we learn that ${\bO}=(O_0,\,O_1)^T$ is a rank-2 multiplet with weight $\Delta=1$ and vanishing boost charge ${\bf\xi}=0$.

In addition, there exist ``vertex operators" in this free BMS scalar model, which are operators of the form
\begin{equation}\label{valpha}
V_\alpha (x,y)\equiv:e^{\alpha \phi(x,y)}:
\end{equation}
From their OPEs with the stress tensor,
\eq{
T(x',y')V_\alpha(x,y)&=\frac{\partial_y V_\alpha}{x'-x} - \frac{(y'-y)\p_y V_\alpha}{(x'-x)^2}+\frac{\alpha^2(y'-y) V_\alpha}{(x'-x)^3},
\\
M(x',y')V_\alpha(x,y)&=\frac{\partial_y V_\alpha}{x'-x} + \frac{-\frac{\alpha^2}{2} V_\alpha}{(x'-x)^2}\,.}
we can read that the vertex operators are singlet primary operators with $\Delta=0$ and $\xi= -\alpha^2/2$.  It is interesting to note that $\alpha$ can be either real or purely imaginary, since there are no constraints on the boost charges other than reality. This is different from the usual case of relativistic theory for free scalars which are 2d CFTs, where unitarity requires positive conformal weights hence purely imaginary $\alpha$'s. 

To summarize, we find that $(O_0,O_1)$ is a rank-2 primary multiplet with weight $\Delta=1$ and boost charge $\bxi=\left(\begin{matrix}
   0&0  \\
   1&0\end{matrix}\right)$. Additionally, the vertex operator $V_{\alpha}$ with $\alpha\in \R \cup i\R$ is a singlet primary operator with $\Delta=0$ and $\xi=-\alpha^2/2$.

\subsubsection{Operator basis}

Given the state space as described by \eqref{basis}, we can find the basis of local operators via the state operator correspondence
 \eqref{operator-state}.

Let us first look at the states that correspond to the Vertex operators $V_\alpha$'s, which by definition are highest weight states carrying exactly the same quantum numbers as the zero mode state $|\alpha\rangle$'s. Therefore we identify
\begin{equation}
|\alpha\rangle= \lim_{x\rightarrow0,y\rightarrow0}V_{\alpha}(x,y)|0\rangle
\end{equation}

Now let us consider the weight 1 primary operators  $O_1=i\partial_x\phi,\, O_0=i \partial_y\phi$ as defined in \eqref{dphi}. One finds
\eq{\lim_{x\rightarrow0,y\rightarrow0} O_1|0\rangle&=i A_{-1}|0\rangle,\\
\lim_{x\rightarrow0,y\rightarrow0}O_0|0\rangle&= i B_{-1}|0\rangle.
}
Their descendants are then
\eq{
 \p_k O_1|0\rangle&=i\partial_x^{k+1}\phi(x,y)|0\rangle= i(k+1)!A_{-k-1}|0\rangle,\\
\p_k O_0|0\rangle&=i\partial_x^{k}\p_y\phi(x,y)|0\rangle= i(k+1)!B_{-k-1}|0\rangle.
}
That is,  the states with a single creation operator $A_{-k}$ correspond to the operator $\partial_x^{k}\phi(x,y)$, and the states with single $B_{-k}$ correspond to the operator $\partial_x^{k-1}\p_y\phi(x,y)$. This relation also works for the composite states, namely
 \eq{\label{composite}
:\partial_x^{k_1}\phi\,  \partial_x^{k_2}\phi\,\,\cdots\,\, \partial_x^{l_1-1}\p_y\phi\,  \partial_x^{l_2-1}\p_y\phi:
 \,\,\sim\,\,  A_{-k_1}\,A_{-k_2}\,\cdots \,B_{-l_1}\,B_{-l_2}\cdots}

Putting all the above together, we learn that the state space are generated by acting operators of the form \eqref{composite} on the zero mode states $|\alpha\rangle$'s, the latter of which correspond to vertex operators.
Therefore we conclude that in the BMS free scalar model, there exists a complete basis of local operators,  \eq{\label{operatorbasis}&\{ \, V_\alpha, \,:\partial_x^{k_1}\p^{\delta_1}_y\phi\cdots \partial_x^{k_n}\p_y^{\delta_n}\phi \,e^{\alpha \phi}:\,\}\\
& n\in \Z^+, \, k_n+\delta_n\ge 1,\, k_n,\in  \N, \, \delta_n=0,1,\,\alpha\in \R\cup i\R.\nonumber
}
which, when inserted at the origin, give all states in the state space spanned by \eqref{basis}. As a special case, the identity operator corresponds to $I=V_0$. The only fundamental primary operators are $\{\partial_x\phi,\partial_y\phi,V_{\alpha}=e^{\alpha\phi},\alpha\in \R\cup i\R\}$.

\subsubsection{Correlation functions}

Correlation functions can be obtained from the OPEs. Let us first consider the rank$-2$ multiplet  \eqref{dphi}.
The two-point functions are given by,
\eq{\langle O_0(x_1,y_1)O_0(x_2,y_2) \rangle&=0
\\ \langle O_0(x_1,y_1)O_1(x_2,y_2) \rangle&=\frac{1}{x_{12}^2}
\\
\langle O_1(x_1,y_1)O_1(x_2,y_2) \rangle&=-\frac{2y_{12}}{x_{12}^3}
}
where $x_{12}=x_1-x_2$, $y_{12}=y_1-y_2$. The two-point functions above agree with the general result for a $\xi=0$ rank-2 multiplet \eqref{xi02pt}.
All three-point functions within the multiplet vanish, namely,
\eq{\langle O_a(x_1,y_1)\,O_b(x_2,y_2) \, O_c(x_3,y_3)\rangle=0  }

Next, the vertex operator $V_\alpha$ \eqref{valpha} satisfies the following OPE,
\begin{equation}\label{vvope}
V_\alpha(x',y') V_\beta(x,y) \sim e^{-\alpha \beta \frac{y'-y}{x'-x}} ~V_{\alpha+\beta},
\end{equation}
 which implies the correlation functions,
\begin{equation}
\langle V_{\alpha}(x_1,y_1)V_{\beta}(x_2,y_2) \rangle= \begin{cases}
  e^{\alpha^2\frac{y_1-y_2}{x_1-x_2}}, &  \alpha+\beta=0 \\
  0, & \alpha+\beta\neq0.
\end{cases}  .
\end{equation}
Using the state operator correspondence, this implies that zero mode background satisfy the following orthonormal condition\begin{equation}
\langle \alpha'|\alpha\rangle=\delta_{\alpha^{\prime},-\alpha}
\end{equation}
More generally, we have
\begin{equation}
\langle \prod_{k=1}^nV_{\alpha_k}(x_k,y_k)\rangle=\exp\{\sum_{i<j}^n(-\alpha_i\alpha_j)\frac{y_i-y_j}{x_i-x_j}\},
\end{equation}
which do not vanish only when the following condition is obeyed,
\begin{equation}
\sum_{k}\alpha_k=0\label{chargecon}
\end{equation}
The above condition can also be understood from the charge conservation of the internal $U(1)$ symmetry \eqref{pl_in_char} . The vacuum is charge neutral as $J_0|0\rangle=B_0|0\rangle=0$, while the vertex operator $V_\alpha$ carries global $U(1)$ charge $\alpha$,
\begin{equation}
[J_0,V_\alpha]=\alpha V_\alpha\,.
\end{equation}
Therefore the condition \eqref{chargecon} is just the condition for charge conservation. Note that the multiplet $(O_0,\,O_1)^T$ is already charge neutral under the global $U(1)$ symmetry, so there are no additional constraints for correlations among these operators.

Finally, let us consider the OPEs between the $(O_0,\,O_1)^T$ multiplet and the vertex operators,
\eq{O_0(x',y')V_\alpha(x,y)& \sim - \frac{i\alpha}{x'-x} V_\alpha(x,y),\\
O_1(x',y')V_\alpha(x,y)& \sim  \frac{i\alpha(y'-y)}{(x'-x)^2} V_\alpha(x,y),
}
which means that the two-point functions between them always vanish. Using Wick's theorem, we find the non-vanishing three-point functions are
\eq{
\langle O_0(x_1)V_\alpha(x_2,y_2)V_{-\alpha}(x_3,y_3)&=\frac{-i\alpha}{x_{12}}e^{\alpha^2\frac{y_{23}}{x_{23}}}+\frac{i\alpha}{x_{13}}e^{\alpha^2\frac{y_{23}}{x_{23}}},\\
\langle O_1(x_1,y_1)V_\alpha(x_2,y_2)V_{-\alpha}(x_3,y_3)&=\frac{i\alpha \,y_{12}}{x_{12}^2}e^{\alpha^2\frac{y_{23}}{x_{23}}}-\frac{i\alpha\, y_{13}}{x_{13}^2}e^{\alpha^2\frac{y_{23}}{x_{23}}}.
}

We end this section with the following concluding remarks,
\begin{itemize}
  \item The quantum theory of the free scalar BMSFT \eqref{action} depends on the choice of the vacuum.  We find a self-consistent highest weight vacuum, where the free BMS scalar has the central charges $c_L=2$ and $c_M=0$.
  \item We calculate the correlators in the highest weight vacuum. The primary operators consist of $\{I, ~ \bO=(O_0=i\p_y \phi, \, O_1=i\p_x \phi), ~ V_\alpha=:e^{\alpha \phi}: | ~\alpha \in \R\cup i\R\}$, where $\bO$ is a primary multiplet with $\Delta=1,\, \xi=0$ and $V_\alpha$'s are vertex operators.
  \item In 
  the context of tensionless string theory, different types of vacua have been discussed,
  including the so-called induced vacuum, flipped vacuum and oscillator vacuum. The induced vacuum is related to our discussion in section 3.2.1 with choice I and II, a detailed discussion of which will be postponed to appendix B.
The so-called flipped vacuum in \cite{Arjun20} is similar to our highest weight vacuum, \eqref{vac1.1} but without requiring $B_0|0\rangle=0,$ and hence is not invariant under the action of $L_{-1}$ and $M_{-1}$.
 The oscillator vacuum is not invariant under the global subgroup of the BMS group either. Thus our highest weight vacuum provides a new starting point for the study of the free scalar BMSFT \eqref{action} as a quantum theory.
  \end{itemize}

\section{The enlarged BMS module}

In the last section we found a basis of the state space \eqref{basis} in terms of annihilation and creation operators $A_{n},\,B_{n}$, and a basis of the local operators \eqref{operatorbasis} in terms of the composite operators constructed from $O_0,\, O_1$.
In this section, we will see how to organize the states and local operators in terms of BMS modules.
Because this model has $c_M=0$, novel features appear, and it turns out the states have to be organized into an enlarged BMS module, which is similar to the so-called staggered module of logarithmic CFTs \cite{Gaberdiel:1996kx, Rohsiepe:1996qj, Gaberdiel:2001tr, Kytola:2009ax, Cardy:2013rqg, Creutzig:2013hma}.

\subsection{Truncation at $c_M=0$? }
In this subsection we revisit the general analysis of BMSFTs with $c_M=0$ \cite{2010}, which states that the BMS module has a truncation as a Virasoro module. We will show that this statement is true provided that there are no extra quasi-primary operators with $\Delta=2$ other than $T$ and $M$. In this case, the theory does not allow multiplets either.

From the OPE of the stress tensor in a generic BMS field theory \eqref{BMSope}, we learn that the stress tensor $T$ and $M$ form a rank-2 multiplet $\bO=\{2M,T\}^T$ with conformal weight $\Delta=\left(\begin{matrix}
   2 &  \\
   & 2\end{matrix}\right)$, and boost charge $\bxi=\left(\begin{matrix}
   0&0  \\
   1&0\end{matrix}\right)$.
An interesting special case is when $c_M=0$, which actually occurs in the free BMS scalar.
In this case,
the state $M_{-2}|0\rangle$ has vanishing inner product with both itself and the state $L_{-2}|0\rangle$.
If we assume that there are no other level 2 states in the vacuum module, $M_{-2}|0\rangle$ will be a null state as it is orthogonal to all states.
By considering the inner products of the higher descendant states, one can similarly arrive at the conclusion that $M_{-n}^i\cdots|0\rangle,\,n>0$ are all null states. Then the vacuum is invariant under the action of all the $M_n$'s for arbitrary integer $n$.
This leads to further constrains on the two point functions,
\begin{equation}
\langle M_n|\bO_i(x_1,y_1)\bO_j(x_2,y_2)|0\rangle=\langle0|M_n\bO_i(x_1,y_1)\bO_j(x_2,y_2)|0\rangle=0,\ \ \forall n\in \Z,
\end{equation}
where we have used the fact that $M$ is the top component of the rank two multiplet so that the out state $\langle M_n|$ is $\langle 0|M_n$ according to the definition \eqref{outstate}.
Using the Ward identity, the above condition leads to the following differential equations,
\eq{
&\langle0|M(x)\bO_i(x_1,y_1)\bO_j(x_2,y_2)|0\rangle\nonumber\\
=&\sum_{k=1,2}(\frac{\partial_{y_k}}{x-x_k}+\frac{\bxi_k}{(x-x_k)^2})\langle0|\bO_i(x_1,y_1)\bO_j(x_2,y_2)|0\rangle=0.\label{nullstate}
}

Two point functions have to satisfy \eqref{nullstate} in addition to the six conditions coming from the global symmetries which leave the vacuum invariant.
Plugging the solution \eqref{2pt} into \eqref{nullstate}, one can check that the allowed solutions have to be $y-$independent and meanwhile have $\xi=0$.
On the other hand, according to the discussion in section 2.3.2, a multiplet with rank $r>1$, there  always exists $0\le a,\,b<r$  satisfying $q=a+b+1-r>0$,  such that $\langle O_a O_b\rangle$ has $y-$dependence even if $\xi=0$ \eqref{2pt}.
Therefore the existence of multiplets is not compatible with \eqref{nullstate}, and such a theory only admits singlets with $\xi=0$.

From the above argument, one may draw the conclusion that the highest weight representation of the BMS algebra has a truncation to the one of the Virasoro algebra \cite{2010}, with no appearance of multiplets.
In the free scalar model, however, we have  explicitly constructed a  multiplet with $\xi=0$.
To understand the apparent discrepancy, let us recall that in the general analysis above, we have assumed that there are no other states in the vacuum module at level two other than $L_{-2}|0\rangle$ and $M_{-2}|0\rangle$. On the other hand, in the free scalar model, there exists a new quasi-primary which is not orthogonal to $M_{-2}|0\rangle$, and hence the aforementioned  truncation does not happen.
We will illustrate this point in more detail in the following subsection.

\subsection{Enlarged BMS module in the free scalar model}
In the general analysis above, we have assumed that the only weight 2 quasi-primary operators in the vacuum module are the stress tensor $T$ and $M$. However, in the free BMS scalar model, we find that there is another weight 2 state $-A_{-1}A_{-1}|0\rangle$, which corresponds to the operator
\begin{equation}\label{defK}
K\equiv - \frac{1}{2}:\partial_x\phi\partial_x\phi:
\end{equation}
The existence of \eqref{defK} violates the assumption in the subsection above, so that the obstruction \eqref{nullstate} of having highest weight multiplets disappears, and the truncation to the Virasoro module will not happen. In fact,  the new operator \eqref{defK}
will enlarge the highest weight module in our free scalar model. To see this explicitly, we first calculate the OPEs using the $\phi\phi$ OPE \eqref{phiphi}, and the results are as follows,
\begin{align}\label{triplet}
M(x',y')K(x,y) &\sim \frac{1}{2\left(x'-x\right)^4} + \frac{T}{\left(x'-x\right)^2} + \frac{\partial_{y}K}{\left(x'-x\right)},\\
T(x',y')K(x,y)&\sim \frac{-2 \left(y'-y\right)}{\left(x'-x\right)^5} - \frac{2 \left(y'-y\right)T}{\left(x'-x\right)^3}  + \frac{2K}{\left(x'-x\right)^2} +  \frac{\partial_{x}K}{\left(x'-x\right)}- \frac{\left(y'-y\right)\partial_{y}K}{\left(x'-x\right)^2}.\nonumber
\end{align}
Comparing \eqref{TMope} and \eqref{triplet} with the defining properties of quasi-primary multiplets \eqref{quasiOPE},
we find that the operators $\bT=\{2M,T,K\}$ actually form a rank-3 quasi-primary multiplet with weight and charge,
\begin{equation}
\Delta=\left(\begin{matrix}
   2 & & \\
   & 2 & \\
   & & 2 \end{matrix}\right),\ \ \
\bxi=\left(\begin{matrix}
   0 &0 &0 \\
   1& 0 &0 \\
   0& 1& 0 \end{matrix}\right).
\end{equation}
One can also explicitly calculate the inner products between states which correspond to different components of $\bT$, and verify that they indeed satisfy \eqref{norm}, namely
\begin{equation}
\langle
\bT|\bT\rangle_{ab}=\delta_{a+b,2},
  \end{equation}
which  clearly shows that $M_{-2}|0\rangle$ is not a null state.
The vacuum is still invariant under the global part of the BMS group generated by $\{L_{0,\pm1},M_{0,\pm1}\}$, but the action of other $M_n$'s provides no further constraints on the correlation functions. In this case, the representation of the BMS algebra does not truncate to that of a Virasoro algebra as was argued in \cite{2010}. Instead, the representation of the BMS algebra is enlarged to the so-called staggered module which we will describe below.

For completeness, we also provide the OPE between $K$ and other primary operators here,
\begin{align}
K(x',y')O_0(x,y) &\sim \frac{O_1}{\left(x'-x\right)^2} + \frac{\partial_xO_1}{\left(x'-x\right)} + \frac{\left(y'-y\right)\partial_xO_0}{\left(x'-x\right)^2}, \\
K(x',y')O_1(x,y) &\sim \frac{-2\left(y'-y\right)O_1}{\left(x'-x\right)^3} - \frac{2\left(y'-y\right)\partial_xO_1}{\left(x'-x\right)^2} - \frac{2\left(y'-y\right)^2\partial_xO_0}{\left(x'-x\right)^3},\\
K(x',y')V_\alpha(x,y) &\sim - \frac{\alpha^2\left(y'-y\right)^2V_{\alpha}}{2\left(x'-x\right)^4}-\frac{\left(y'-y\right)\partial_xV_{\alpha}}{\left(x'-x\right)^2}.
\end{align}

\subsubsection*{The representation}
In section 3 we discussed the operator basis  where the primary operators were found to be $I,\,\bO$ and $V_\alpha$'s, and other operators should be organized into one of the three families. We have just learnt that $(2M,\,T,\,K)$ form a quasi-primary triplet, which means that $K$ should belong to the vacuum module \footnote{We will not distinguish the notion of states and operators, and hence of BMS modules and operator families in this section.}. However, in the usual BMS highest weight representation, the vacuum module only contains the composite operators of $T$ and $M$.
To accommodate the new operator $K$, the ordinary highest weight module should be enlarged to the so-called staggered BMS module, which is an indecomposable representation of the BMS algebra, defined as the semi-direct product of two ordinary highest weight representations. While it is interesting to analyze this enlarged module at $c_M=0$ from a more general point of view, in this paper we restrict our discussion to the free scalar model, and illustrate how states are organized into the new module below.

\begin{figure}[htbp]
\includegraphics[width=3in]{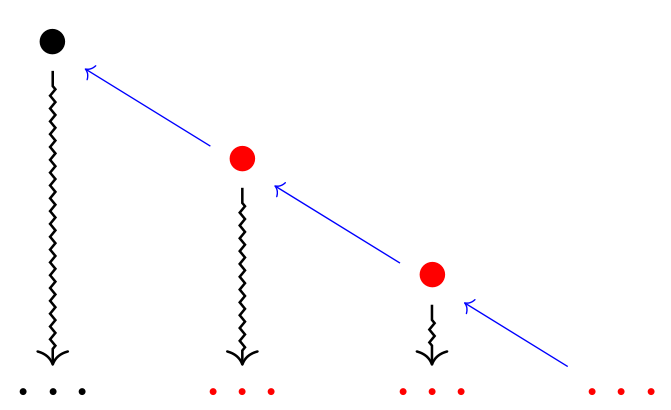}
\caption{Staggered module} \label{staggered}
\end{figure}

The staggered BMS module can be constructed as follows, starting from a primary state $O$, or more generally a primary multiplet $\bO$, we can first construct the ordinary BMS module by applying $L_{-n},\,M_{-n}$ with $n>0$ successively. To enlarge the module, we add one more state which corresponds to the composite operator $:K\bO:$, and construct its BMS descendants by acting with $L_{-n},\,M_{-n}$ successively. We will refer to states descended from $\bO$ (including $\bO$ itself) as the main branch, and states from $:K\bO:$ as the first side branch.
Similarly, the composite operators $:KK\bO:, \,:KKK\bO:$, etc, and their BMS descendants form the second side branch, the third side branch, etc.
As a result,  the primary operators $\bO$ and the composite operators $:K^n\bO:$ are all seeds of the enlarged module, from which we can build infinite many branches of states by acting with the raising operators of the BMS algebra.
In order to form a single BMS module instead of separate modules, the branches must be bonded together. As we will see explicitly momentarily, it is $M_{0}$ that sews the states in different branches together.
Consequently, states at each level will be grouped into several multiplets, and states within each multiplet are related by the action of $M_0$.
Interestingly, if we apply the lowering operators $L_n,\,M_n, \,n\ge 1$ to states in the side branches, we may obtain states in the main branch, whereas states in the main branch will only flow within the main branch.
Key features of the staggered module include:
 \begin{itemize}
 \item Removing the side branches, we are left with the main branch, which is the usual highest weight module.
 \item The first side branch can be viewed as a highest weight module if we mod out the main branch; the second side branch can be viewed as a highest weight module if we mod out  both the main branch and first side branch;
similarly, the $(n+1)$-th side branch can be viewed as a highest weight module if we mod out the first to the $n$-th side branches as well as the main branch.
 \item The seed of a side branch can be mapped to the seed of the main branch by lowering operators, whereas there are no raising operators to map the seed of the main branch to seeds of the side branches.
 \end{itemize}
The full structure of the staggered module is schematically depicted in Fig. \ref{staggered}, where the black dot represents the primary $\bO$, red dot represents the new seeds $:K\bO:,\,:KK\bO:,\dots$, each vertical squiggly arrow represents a branch descended from a seed operator, and the blue arrows represents the action of $M_2$ which maps the seeds from different branches.
This structure is very similar to logarithmic CFTs with $c=0$ \cite{Gaberdiel:1996kx, Rohsiepe:1996qj, Gaberdiel:2001tr, Cardy:2013rqg, Creutzig:2013hma, Kytola:2009ax}, where the Virasoro stress tensor is accompanied by a logarithmic partner and the Virasoro module is also enlarged to a staggered module.

In the following, we use diagrams to illustrate different modules in the free scalar model, where we use solid dots for states, downward arrows for raising operators $L_{-n},M_{-n}$ with $n>0$, horizontal  arrows for $M_0$, and upward arrows for lowering operators $L_{n},M_{n}$ with $n>0$.  We also use different colors to distinguish different generators. The Virasoro generators are in black and the supertraslations generators $M_n$ are all in blue. States that can be viewed as a Virasoro descendant of the primary are colored in black, states that are BMS descendants but not Virasoro descendants are colored in blue, and new states related to the operator $K$ are all in red.
At each level, we always put the states with more $B_{-n}$'s to left.
As  the action of $M_{n}$ decreases the number of $A_{-n}$'s and increases the number of $B_{-n}$'s,  all the blue arrows representing  $M_n$ should always point left.  In our convention, states that are linked by horizontal blue lines are within a multiplet.

\subsubsection*{The vacuum module}
\begin{figure}[H]\centering
\includegraphics[width=3in]{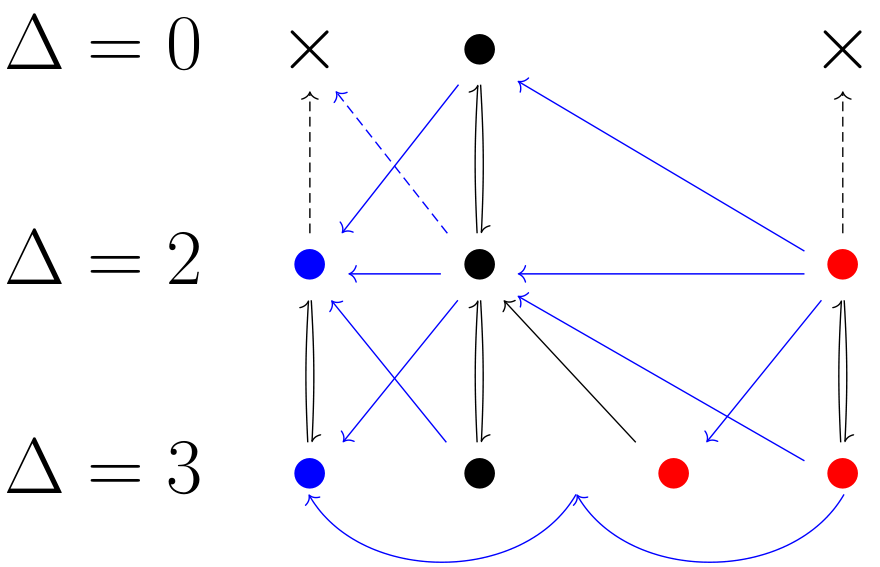}
\caption{
The vacuum module:
the vacuum state with $\Delta=0$;\\
states with $\Delta=2$ form a quasi-primary multiplet, from left to right:
$ |M\rangle,\,  |T\rangle, \, |K\rangle$; \\
states with $\Delta=3$ from left to right:
$L_{-1}|M\rangle,\,  L_{-1}|T\rangle,\, M_{-1}|K\rangle, \,L_{-1} |K\rangle$.\\
The four states at level 3 split into two multiplets: a singlet $L_{-1}|T\rangle-3M_{-1}|K\rangle$, and a triplet consisting of $L_{-1}|M\rangle$, $L_{-1}|T\rangle+M_{-1}|K\rangle$ and $ L_{-1} |K\rangle$.
}
\label{vacuumdig}\end{figure}

Let us first consider the vacuum module up to states with $\Delta=3$, as depicted in Fig. \ref{vacuumdig}. Up to this level, the vacuum module only contains two seeds, the vacuum at level zero and the quasi-primary state $|K\rangle$ at level two. Other states in the module can be generated from the seeds by the raising operators $L_{-n}$ and $M_{-n}$ with $n>0$ represented by downward arrows. Now let us comment on the states at each level. There is a unique state at level zero, the vacuum state, which is represented by the black dot in the middle of the first line. There are no states at level one, as both $L_{-1} $ and $M_{-1}$ annihilate the vacuum. The three states at level 2 form a rank 3 quasi-primary multiplet, of which two states $B_{-1}^2|0\rangle\sim |M\rangle$ and $A_{-1}B_{-1}|0\rangle\sim |L\rangle$ are in the main branch, and the new state $\,A_{-1}^2|0\rangle\sim| K\rangle$, represented by the red dot on the right end, seeds the first side branch. The four states at level 3 split into two multiplets: a singlet $L_{-1}|T\rangle-3M_{-1}|K\rangle$, and a triplet consisting of $L_{-1}|M\rangle$, $L_{-1}|T\rangle+M_{-1}|K\rangle$ and $ L_{-1} |K\rangle$. Note that we have omitted  the links representing the direct action of $L_{-3},\,M_{-3}$ in the figure, since they can  respectively be expressed in terms of  $L_{-1}L_{-2}$ and $L_{-1}M_{-2}$.

Finally, applying lowering operators, which run upwards, will add further links between states. For example, there is a blue arrow pointing to  northwest on the lower left part of Fig. \ref{vacuumdig}, illustrating the relation $M_{1}(L_{-1}|T\rangle)\sim |M\rangle$.
An interesting feature is that usually arrows for $L_n$'s run double directions, while those for $M_n$'s do not. For example, acting $M_{2}$ on the new seed  $|K\rangle$ will get the vacuum, but there is no raising operator that maps the vacuum to the $K$ state. In addition, the $K$ state is annihilated by the Virasoro lowering modes $L_n,\,n>1$, represented by dashed lines.  For simplicity, we have omitted all null states except for those at level  zero which are represented by the symbol~$\times$. From Figure \ref{vacuumdig}, it is clear that the vacuum module does not only contain the Virasoro descendants which are represented by black dots, but also contain the $M$-descendants colored in blue, and the novel $K$-states colored in red. Thus, instead of a truncation, we have an enlargement of the BMS highest weight representation.

\subsubsection*{The $\bO$ module}
\begin{figure}[h!]

\begin{floatrow}
    \centering
    \ffigbox{
    \includegraphics[width=0.73\linewidth]{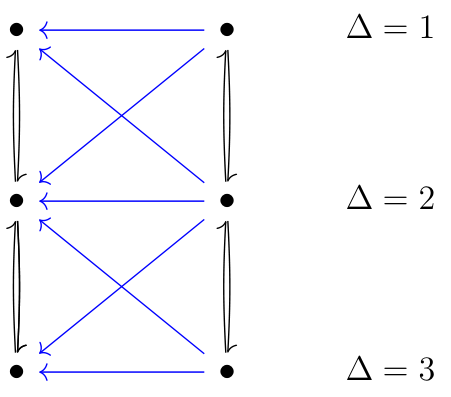}
}{\caption{
$\bO$ module, part I:\\
 states with $\Delta=1$:  $|O_0\rangle,\, |O_1\rangle$;\\ states with $\Delta=2$: $L_{-1}|O_0\rangle,\, L_{-1}|O_1\rangle$;\\ states with $\Delta=3$: $L^2_{-1}|O_0\rangle,\, L^2_{-1}|O_1\rangle$.}\label{OI}}
    \ffigbox{
       \includegraphics[width=1\linewidth]{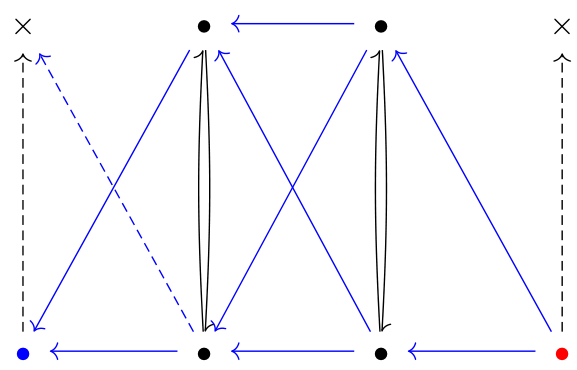}
}{\caption{
$\bO$ module, part II :\\
 states with $\Delta=1$:  $|O_0\rangle,\, |O_1\rangle$;\\
 states with $\Delta=3$:\\
$M_{-2}|O_0\rangle,\,L_{-2}|O_0\rangle,\, L_{-2}|O_1\rangle,\,  |KO_1\rangle$.}\label{OII}}
    \end{floatrow}
\end{figure}

Now let us consider the module seeded by the primary multiplet $\bO=(O_0,\,O_1)^T$, which has conformal weight $\Delta=1$.
The $\bO$ module also shares the key features of a staggered module, although it is more complicated than the vacuum module.
To make the picture clear, we split the $\bO$ module into two figures.
Figure \ref{OI} contains the primary multiplet $\bO$ and
their $SL(2,\,\mathbb R)$-descendants, $L_{-1}\bO $ with $\Delta=2$ and $L_{-1}^2 \bO$ with $\Delta=3$.
States at each level form a multiplet as indicated by the horizontal blue lines.
Note that Figure \ref{OI} does not contain new states corresponding to the presence of the operator $K$. In Figure \ref{OII}, the two states with $\Delta=1$ are also the primary multiplet $\bO$, and the four states with $\Delta=3$, which from left to right are $M_{-2}|O_0\rangle,\,L_{-2}|O_0\rangle,\, L_{-2}|O_1\rangle$ and $|KO_1\rangle$, form a rank 4 multiplet.  Again, the new state  $|KO_1\rangle$ colored in red provides a seed for the enlargement of the representation.

Putting the vacuum module and the $\bO$ module together, we summarize, up to $\Delta=3$, the number of states, quasi-primary states, primary states  and also the organization of the multiplets in the table below. In the last line, we use numbers in bold font to indicate the rank of the multiplets. For example, ${\bf 3}+{\bf 2}$ means that the 5 states with $\Delta=2$ split into a multiplet of rank $3$ and a multiplet of rank $2$.
\begin{figure}[H]\begin{center}
    \begin{tabular}{|c|c|c|c|c|}
\hline
conformal weight  &  $\Delta=0$ &  $\Delta=1$ & $\Delta= 2$ & $\Delta= 3$  \\
\hline
 \# of states &  1 &  2 &  5 &  10  \\
\hline
 \# of quasi-primaries  &  1 &  2 &  3 &  4  \\
\hline
 \# of primaries &  1 &  2 &  0 &  0 \\
\hline
multiplets &  {\bf 1} &  {\bf 2} &  {\bf 3}+{\bf 2} &  {\bf 3}+{\bf 1}+{\bf 4}+{\bf 2}\\
\hline
\end{tabular}\\
\end{center}
\caption{
States up to $\Delta=3$
}
\end{figure}

To recapitulate, we have learnt that BMS field theories with $c_M=0$ are special and subtle.  In a general BMS field theory, the stress tensors $T,\,M$ are in a multiplet. Depending on the details of the theories, there are two possibilities for the representation theories as following.
\begin{itemize}
  \item
  If $T,\,M$ are the only quasi-primary operators with $\Delta=2$, then the highest weight representation of the BMS algebra reduces to that of the Virasoro algebra. In particular, BMS multiplets should not appear, and all states should have vanishing boost charge $\xi=0$.
  \item   If there are other  quasi-primary fields with $\Delta=2$, so that the multiplet containing $T,\,M$ is enlarged, then the highest weight representation of the BMS algebra is also enlarged. BMS primary multiplets can appear, and the truncation does not happen.
 \end{itemize}
For the BMS free scalar model with $c_M=0$, we indeed find an extra quasi-primary operator $K$, which provides a seed to enlarge the ordinary highest weight module. We have also explicitly found a rank-2 primary multiplet $\bO$.
It would be interesting to study general BMS field theories with $c_M=0$.
In particular, the associativity of the operator algebra can be used to constrain the stress tensor multiplet, which can help us to systematically classify this class of theories. We leave this to further work.

\section{Ultra-relativistic limit from CFT$_2$}
So far we have been discussing the free scalar model as an intrinsic BMSFT. Alternatively, it is also useful to
make connections to relativistic CFT$_2$s. Starting from the highest weight representation of a CFT$_2$, we can construct the so-called flipped representation as we will describe momentarily. Then BMSFT in the highest representation can be obtained as the Ultra-relativistic limit of CFT$_2$ in the flipped representation.
In this section, we will first revisit the flipping+UR limit of the CFT$_2$ on the cylinder discussed in \cite{ 2010, Bagchi:2019unf, Bagchi:2017cpu, Bagchi:2014iea}, and point out a subtlety in the limit on the plane.  Then we will discuss the free BMS scalar model as  a flipping+UR limit of a free scalar CFT. Starting from a CFT$_2$, one can also take the non-relativistic (NR) limit to obtain a Galilean conformal theory, which we will discuss in appendix C. For the BMS algebra and field theories as the UR limit of CFT$_2$s, please also see \cite{Barnich:2012aw,Barnich:2012rz}.
\subsection{General discussion on the UR limit}
\subsubsection{UR limit on the cylinder}
Consider a CFT$_2$ on the cylinder parameterized by $\sigma, \,t$ with the periodicity condition
\begin{equation}
\sigma\sim\sigma+2\pi.
\end{equation}
The infinitesimal conformal transformations are generated by
\begin{equation}\label{cylinderlim}
l^+_{n}={i\over2} e^{i(\sigma+t)n}(\partial_\sigma+\p_t),\ \ l^-_{n}=-{i\over2}e^{-i(\sigma-t)n}(\partial_\sigma-\p_t).
\end{equation}
Conformal transformations are implemented in CFT$_2$ by the Virasoro generators $L^+_{n},L^-_{n}$  which form
two copies of the Virasoro algebra \eq{\label{CFT2alg}
\left[L^+_{n},\,L^+_{m}\right] &= (n-m)L^+_{m+n}+\frac{c}{12} n(n^2-1)\delta_{m+n,0}\nonumber\\
  \left[L^-_{n},\,L^-_{m}\right] &= (n-m)L^-_{m+n}+\frac{\bar{c}}{12}  n(n^2-1)\delta_{m+n,0},\\
   \left[L^+_{n},\,L^-_{m}\right] &= 0\nonumber
}
The UR limit on the cylinder is defined as
 \begin{equation}\label{URlim}
t =\epsilon\tau,\ \quad \epsilon\rightarrow 0 \end{equation}
 so that the speed of light goes to zero, which is the reason why this limit is called the ultra-relativistic limit.
 Under this limit the conformal transformations become BMS transformations generated by
 \begin{equation}
l_n=l^+_{n}-l^-_{-n}=ie^{in\sigma}(\partial_\sigma+in\tau\partial_\tau),\ \ \ m_n=\epsilon(l^+_{n}+l^-_{-n})=ie^{in\sigma}\partial_\tau  \label{bms_gen_cyl}
\end{equation}
 Correspondingly, the Virasoro algebra \eqref{CFT2alg} becomes the BMS algebra \eqref{BMSalg}
 via a Wigner-In\"{o}n\"{u} contraction \cite{InonuWigner},
\eq{\ \ L_n=L^+_{n}-L^-_{-n},\ \ M_n=\epsilon(L^+_{n}+L^-_{-n}).\label{contraction}
}
with the central charges related by,
\begin{equation}\label{57URc}
c_L=c-\bar{c},\ \ c_M=\epsilon(c+\bar{c}).
\end{equation}

\subsubsection{UR limit on the plane}
It is also useful to spell out the UR limit on the plane, which we use extensively in the free scalar model. This limit turns out to be more subtle, and our discussion below is different from previous discussions in the literature \cite{2009, 2010,Bagchi:2016geg}.
Before the limit, the map from cylinder parameterized by $(\sigma, \,t)$ to plane parameterized by $(z,\bar{z})$ is given by,
\begin{equation}
z=e^{i(\sigma+t)},\ \ \bar{z}=e^{-i(\sigma-t)}. \label{cft_map}
\end{equation}
and the CFT$_2$ generators become,
\begin{equation}\label{c511g}
l^+_{n}=-z^{n+1}\partial_z,\ \ l^-_{n}=-\bar{z}^{n+1}\partial_{\bar{z}}.
\end{equation}
Our goal is to find a limit of the coordinates which will give well-defined BMS generators and thus a BMS algebra under the contraction \eqref{contraction}. One naive guess is to take a limit similar to the one on the cylinder \eqref{cylinderlim}
\begin{equation}\label{naiveplane}
z=\epsilon y+x,\ \ \bar{z}=-\epsilon y+x,\ \ \epsilon\rightarrow0
\end{equation}
As discussed in \cite{2010}, an analogy of \eqref{naiveplane} does work in the non-relativistic case and it leads to a well-defined GCA, which we review in the appendix C.
However, under the naive limit \eqref{naiveplane}, the generators become
\eq{ l_n=-\frac{x^{1-n}+x^{1+n}}{2\epsilon}\partial_y,\quad m_n=\frac{x^{1-n}-x^{1+n}}{2}\partial_y}
and hence do not have a well defined UR limit.
To get finite generators, we find that the proper UR limit on the plane should be chosen as
\begin{equation}
z=\epsilon y+x,\ \ \bar{z}^{-1}\equiv \tilde{z}=-\epsilon y+x,\ \ \epsilon\rightarrow0 \label{planeUR}
\end{equation}
Note that in terms of the $(z,\tilde{z})$ coordinates, the CFT$_2$ generators on the plane can be rewritten as
\begin{equation}
-l^-_{-n}\equiv\tilde{l}^-_n=\bar{z}^{-n+1}\partial_{\bar{z}}=-\tilde{z}^{n+1}\partial_{\tilde{z}}.\label{flipz}
\end{equation}
This allows us to define BMS generators similar to \eqref{bms_gen_cyl}
\begin{equation}
l_n=l^+_{n}+\tilde{l}^-_{n}=-x^{n+1}\partial_x-(n+1)x^ny\partial_y,\ \ \ m_n=\epsilon(l^+_{n}-\tilde{l}^-_n)=-x^{n+1}\partial_y
\end{equation}
As a consistency check, one can easily verify that the UR limit on the plane \eqref{planeUR} can also be obtained from
the UR limit on the cylinder \eqref{URlim}
via the BMS plane-to-cylinder map  \eqref{plane2cylinder}.

\subsubsection{Representations}
Starting from a CFT$_2$ with highest weight representations, the UR limit leads to the induced vacuum.
On the other hand, two dimensional conformal algebra allows other representations as well.
In \cite{Bagchi:2017cpu,Bagchi:2019unf}, two types of vacua before the UR limit have been discussed, the highest weight vacuum and the so-called flipped vacuum, which respectively become the induced vacuum and the highest weight vacuum under the UR limit.
We will consider the ``flipped vacuum $\rightarrow$ highest weight vacuum" here, and postpone the discussion on the ``highest weight vacuum $\rightarrow$ induced vacuum" to appendix B.

The flipped representation in CFT$_2$ can be understood as the highest weight representation in the flipped coordinates \eqref{planeUR} before the UR limit, or equivalently, as an automorphism of the right-moving Virasoro algebra,
\begin{equation}
 L^-_n\rightarrow \tilde L_n= -L^-_{-n},\ \ \bar{c}\rightarrow \bar{\tilde c}=-\bar{c},	\label{flip_auto}
\end{equation}
as suggested by \eqref{flipz}.
Starting from an ordinary CFT$_2$ with central charges $c,\bar c$, and the usual highest weight representation with conformal weight $h,\bar h$,  the resulting flipped representation satisfies
\begin{equation}
L^+_0O=hO,\ \ L^-_0O=-\bar{h}O
\end{equation}
\begin{equation}
L^+_{n}O=L^-_{-n}O=0,\ \ n>0
\end{equation}
Under the UR limit \eqref{contraction}, the flipped representation becomes the highest weight representation of the BMS algebra,
\begin{equation}
L_0O=\Delta O,\ \ M_0O=\xi O
\end{equation}
\begin{equation}
L_{n}O=M_{n}O=0,\ \ n>0
\end{equation}
where
\begin{equation}
\Delta=h+\bar{h},\ \ \ \xi=\epsilon(h-\bar{h})
\end{equation}
Note that the above looks very similar to the NR limit \eqref{c11nr}.

\subsection{UR Limit of the relativistic free scalar model}
In this subsection we show that our free scalar BMSFT \eqref{action} can be obtained from the UR limit of a free scalar CFT$_2$.
Consider the free scalar model on the cylinder, 
\eq{\label{rel_scalar_action}
S=\frac{1}{4\pi}\int d\sigma dt\Big((\partial_{t}\Phi)^2-(\partial_{\sigma}\Phi)^2\Big) ,\qquad(\sigma,\,t)\sim (\sigma+2\pi,\,t)
}
Under the UR limit \eqref{URlim} together with the corresponding rescaling of the field,
\begin{equation}
t=\epsilon \tau,\quad
\Phi=\sqrt{\epsilon}\phi,\quad \epsilon\rightarrow0,  \label{model_UR_limit}
\end{equation}
the action \eqref{rel_scalar_action} becomes the BMS scalar action \eqref{action} on the cylinder $(\sigma,\tau)\sim (\sigma+2\pi,\tau)$, which we reproduce here,
\eq{
S=\frac{1}{4\pi}\int d\sigma d\tau (\partial_{\tau}\phi)^2\,.
}
The equation of motion of the relativistic scalar \eqref{rel_scalar_action} can be solved in terms of the mode expansion
\begin{equation}
\Phi=\phi_0+\pi_0t+\frac{i}{\sqrt{2}}\sum_{n\neq0}\frac{1}{n}(a_ne^{-in(\sigma+t)} - \bar{a}_{-n}e^{-in(\sigma-t)}).\label{modes_rel_scalar}
\end{equation}
with the canonical commutation relations \begin{equation}
\label{combefore}[a_n,a_m]=[\bar a_n,\bar a_m]=n\delta_{n+m,0},\ \ [a_n,\bar a_m]=0,\ \ [\phi_0,\pi_0]=i
\end{equation}
Comparing with the mode expansion of the BMS free scalar on the cylinder \eqref{modecylinder}
we obtain the relation between modes before and after the UR limit
\eq{
&A_n=\lim_{\epsilon\to0}\frac{i}{\sqrt{2}n\sqrt{\epsilon}}(a_{n} - \bar{a}_{-n}),~ B_n=\lim_{\epsilon\to0} \frac{-i\sqrt{\epsilon}}{\sqrt{2}}(a_n +\bar a_{-n}),~n\neq 0, \label{modeslimit}\\
&A_0 = \frac{\phi_0}{\sqrt{\epsilon}},~ B_0=-i\sqrt{\epsilon}\pi_0.
}
As a consistency check, one can verify that under such relation the commutation relations before the UR limit \eqref{combefore} indeed become \eqref{c347r} after the limit \eqref{model_UR_limit}. Besides, the central charges in the flipped representation of the relativistic free scalar are $c=1,\bar{c}=-1$, due to \eqref{flip_auto}. After the UR limit, the central charges become $c_L=2,c_M=0$, due to \eqref{57URc}.

This model also enables us to show explicitly why the UR limit on the plane should be \eqref{planeUR}
instead of the naive limit \eqref{naiveplane}. If the latter is taken, the plane mode expansion will contain a zero mode part $- i \log(x) \pi_0$, which does not exist in the BMS mode expansion \eqref{mode}.
This issue, on the other hand, does not appear if we take the proper UR limit \eqref{planeUR}, where the zero mode term in question now becomes $- i {y\over x} \pi_0$ which further becomes the ${y\over x} B_0$ term in the BMS mode expansion under \eqref{modeslimit}. Therefore, the UR limit \eqref{planeUR} is compatible not only with the plane to cylinder map \eqref{plane2cylinder}, but also with the UR limit of the mode expansion.

\subsubsection*{Flipped $\rightarrow$ highest weight representation}
As we discussed in general in subsection 5.1.3, if we take the UR limit of a CFT from the flipped representation, we will obtain a highest weight representation in the BMSFT. Now we want to apply this to our relativistic free scalar model. We will provide evidence that this resulting representation of the BMS free scalar theory is just the representation that we discussed in section 3, by showing that the vacuum obtained from this limit agrees with the one \eqref{phiphi} obtained from intrinsic BMSFT quantization method.

As mentioned earlier, the flipped representation is derived from the usual highest weight representation by the automorphism \eqref{flip_auto} of the right-moving Virasoro. In the free scalar model, this automorphism can be realized at the level of creation and annihilation operators,
\eq{
&\bar{a}_n\to\bar{a}_{-n},  \label{auto_cft_modes}
}
Under this automorphism, the ``ground'' state $|0\rangle$ in the flipped representation is specified by
\eq{\label{f529v}
a_n|0\rangle=\bar{a}_{-n}|0\rangle=0,\ \ n\geq0.
}
Using the relation \eqref{modeslimit} of the modes under the UR limiting procedure, we obtain that the resulting vacuum in the BMS theory satisfies \eq{
A_n |0\rangle &=0,~n>0,\\
B_n |0\rangle &=0,~n\geq 0.
}
We recognize this as the BMS highest weight vacuum defined in \eqref{vac1.1}. Since we have the same vacuum, all the calculations based on the vacuum should be the same.

In particular, the Green's function on the plane in the flipped representation reduces to that of BMSFT \eqref{green}. To do so, we first compute the Green's function in CFT$_2$ by summing over all the modes \eqref{modes_rel_scalar} on the flipped vacuum \eqref{f529v}. Using the commutation relations \eqref{combefore}, this amounts to summing over the $\langle0_F|a_na_{-n}|0_F\rangle$ and $\langle0_F|\bar a_{-n}\bar a_{n}|0_F\rangle\ $ terms with $n>0$, together with a zero-mode term. Further using the map from cylinder to plane \eqref{cft_map},
the Green's function on the flipped vacuum can be written as
\begin{equation}
\langle\Phi(z_1,\bar{z}_1)\Phi(z_2,\bar{z}_2)\rangle_F=-\frac{1}{2}\Big(\log(1-\frac{z_2}{z_1})-\log(1-\frac{\bar{z}_1}{\bar{z}_2})+\log(z_1\bar{z}_1)\Big),
\end{equation}
or more conveniently, in the $\tilde{z}=\bar{z}^{-1}$ coordinate, as
\begin{equation}
\langle\Phi(z_1,\tilde{z}_1)\Phi(z_2,\tilde{z}_2)\rangle_F=-\frac{1}{2}(\log (z_1-z_2)-\log(\tilde{z}_1-\tilde{z}_2)),
\end{equation}
Under the plane UR limit \eqref{planeUR},  the above Green's function indeed becomes that of the BMSFT in the highest weight vacuum \eqref{green}.

We have just shown that the highest weight representation of our free BMS scalar model comes from the UR limit of the flipped representation in the 2d free relativistic massless scalar model.
In the literature, it has also been stated that the UR limit of the highest weight representation in CFT$_2$ leads to the induced representation in BMSFTs. We will discuss this other limit in appendix B.

\section{Torus partition function}
In this section we explicitly calculate the torus partition function for the free scalar BMSFT, and show that it is modular invariant as expected from the general analysis of \cite{Bagchi:2019unf, Bagchi:2012xr, Barnich:2012xq, Song-Wen-Jiang}.

\subsection{Modular invariance of BMSFT}
In this subsection we review the derivation of modular invariance for BMSFTs, which helps us to set up the conventions. Those who are already familiar with this story can skip this subsection.
The torus partition function for BMSFTs has been shown to be modular invariant, by taking a limit of CFT$_2$ in \cite{Bagchi:2012xr, Bagchi:2019unf}, or intrinsically as in \cite{Barnich:2012xq, Song-Wen-Jiang}.
We review the argument of modular invariance for BMSFT, following the intrinsic argument as in Appendix A in \cite{Song-Wen-Jiang}.
We consider a torus which is determined by two identifications on a two dimensional plane \footnote{It is useful to embed $\mathbb R^2$ into $\mathbb C^2$ in the subsequent discussions.},
\eq{
(canonical) \,spatial\, circle:\quad &(\tau,\sigma)\sim(\tau,\sigma+2\pi)\label{canonical}\\
thermal\, circle:\quad &(\tau,\sigma)\sim(\tau-2\pi i b,\sigma-2\pi i a)
}
with $a\in \mathbb R,\, b\in \mathbb R$.
Since the change of orientation of the complexifed $\sigma$ can be realized by the symmetry $(a,b)\rightarrow(-a,-b)$, we will only consider $a>0$ without loss of generality.
The partition function on the above torus is formally a path integral over all fields satisfying boundary conditions specified by the two identifications.
Alternatively, the torus partition function can be written as a trace over the state space which is determined by the {\it spatial circle}, weighted by the evolution along the {\it thermal circle},
\begin{equation}\label{zbth}
Z(a, b)=Tr\,e^{-2\pi a(L_0-\frac{c_L}{24})-2\pi b (M_0-\frac{c_M}{24})}
\end{equation}
where the translational generators are defined on the cylinder with the spatial circle \eqref{canonical}, which we refer to as the canonical circle.
We have also taken the Casimir effect \eqref{cylindershift} into account.

More generally, a torus can be described by the fundamental region on the plane
\begin{equation}
(\tau,\,\sigma)\sim(\tau,\,\sigma)+m\,\vec{\beta}_S +n\,\vec{\beta}_T
\end{equation}
where $m$ and $n$ are integers, so that the torus is completely determined by a pair of vectors $\vec{\beta}_S,\, \vec{\beta}_T$ on the plane.
For instance, the torus \eqref{canonical} has a canonical spatial circle $\vec{\beta}_S=(0,\,2\pi)$, and a thermal circle $\vec{\beta}_T
=(-2\pi i b,\, -2\pi i a)$.
The transformations acting on the plane that leave the torus invariant form the modular group, $SL(2,\,\mathbb{Z})/\mathbb{Z}_2$.  The action of $SL(2,\,\mathbb{Z})$ is given by
\begin{equation}
\left(
\begin{aligned}
a\ \ &b\\
c\ \ &d
\end{aligned}
\right)\left(
\begin{aligned}
\vec{\beta}_S\\
\vec{\beta}_T
\end{aligned}
\right)=
\left(
\begin{aligned}
\vec{\beta}_S'\\
\vec{\beta}_T'\end{aligned}
\right)
\end{equation}
with
\begin{equation}
ad-bc=1, \ \ \ \ a,b,c,d\in \mathbb{Z}.
\end{equation} The reason to mod $Z_2$ is because the simultaneous inversion of all the matrix elements does not change the torus.
The modular group is generated by the $T$ and $S$ transformations, with \eq{
T=\left( \ba{cc}
1&1\\
0&1
\ea
\right),\quad
S=\left( \ba{cc}
0&-1\\
1&0
\ea
\right).}  In particular, the modular $S$ transformation swaps the spatial and thermal circles.
From the path integral point of view, the torus partition function only depends on the torus and hence should be
 invariant under the action of the modular group, namely
\eq{
&Z_{\vec{\beta}_S}(\vec{\beta}_T)\equiv Tr_{\vec{\beta}_S}\,e^{-i  (M_0^{\vec{\beta}_S},\,\, L_0^{\vec{\beta}_S}) \cdot \vec{\beta}_T}=Z_{\vec{\beta}_S'}(\vec{\beta}_T')=Z_{-\vec{\beta}_T}(\vec{\beta}_S)\label{swap}
}
where the trace is taken over the state space, and the translational operators $M_0^{\vec{\beta}_S},\,\, L_0^{\vec{\beta}_S}$ are both defined on the spatial circle specified by $\vec{\beta}_S$, as the subscript and superscript suggest.
Note that the modular group is the isometry group acting on the modular parameters, and hence is independent of the theory. In general, the relation \eqref{swap} is a relation between theories with state spaces defined on different spatial circles.
For two dimensional quantum field theories with enough symmetries, such as  such as $CFT_2$, $BMSFT$, and $WCFT$,
a symmetry transformation can be found to transform the torus such that the spatial circle is transformed back to the original one specified by $\vec{\beta}_S$ again.
For BMSFT on the canonical spatial circle \eqref{canonical}, such a transformation is a BMS symmetry of the form \eqref{BMSfinite}, given by,
\begin{equation} \label{Strans}
f(\sigma)=-{i\over a}\sigma,\ \ g(\sigma)=-\frac{ib}{a^2}\sigma
\end{equation}
under which the identifications $\vec{\beta}'_S=(-2\pi ib',-2\pi ia'), \, \vec{\beta}'_T=(0,\,2\pi)$ after the swapping of the cycles becomes,
\begin{equation}
(\tau,\sigma)\sim(\tau,\sigma-2\pi)\sim(\tau+\frac{2\pi ib}{a^2},\sigma-\frac{2\pi i}{a})
\end{equation}
Finally, using the transformation rules \eqref{translaw} for the finite BMS transformation \eqref{Strans}, one can relate the partition functions before and after the S transformation, and find that it is modular invariant
\eq{
Z(a,b)=Z(\frac{1}{a},-\frac{b}{a^2}).\label{modularinv}
}
Note that the modular group will keep $a>0$.
\subsection{Torus partition function for the free scalar model}
In this subsection we explicitly calculate the torus partition function for the free scalar model. The result depends on the choice of the vacuum. We will perform the calculation in the highest weight vacuum and postpone that of the induced vacuum to appendix B.

\subsubsection*{Intrinsic calculation in the highest weight vacuum}
To calculate the torus partition function \eqref{zbth}, we need to specify both the state space and the explicit definition of the trace.
For the highest weight vacuum, the state space is spanned by \eqref{basis}
\begin{equation}
|\vec{i},\vec{j}; \alpha\rangle=A_{-1}^{i_1}A_{-2}^{i_2}\cdots B_{-1}^{j_1}B_{-2}^{j_2}\cdots|\alpha\rangle.\label{ketbasis}
\end{equation}
Using the conjugation relations \eqref{m34c}, the basis for the out states can be written as
\begin{equation}\label{outbasis}
\langle\vec{i},\vec{j}|=(-1)^j\langle \alpha|\cdots B_{2}^{j_2}B_{1}^{j_1}\cdots A_{2}^{i_2}A_1^{i_1}
\end{equation}
where the overall sign $(-1)^j$ is determined by
\begin{equation}
j\equiv\sum_{k}j_k\,.
\end{equation}
It is not difficult to verify that inner products between states with different zero mode charges are orthogonal with each other, while
the inner products between different states with the same zero mode charge form a non-diagonal matrix $N_\alpha$, namely
\begin{equation}\label{inner}
\langle \vec{i'},\vec{j'};\alpha'|\vec{i},\vec{j};\alpha\rangle=\delta_{\alpha,\alpha'} N_{\alpha;\vec{i}\vec{j},\vec{i'}\vec{j'}}\,,\quad
N_{\alpha;\vec{i}\vec{j},\vec{i'}\vec{j'}}=\delta_{\vec{i'},\vec{j}}\delta_{\vec{j'},\vec{i}}
\end{equation}
The fact that the inner product matrix  between the out-states \eqref{outbasis} and the in-states \eqref{ketbasis} is not diagonal requires a more careful definition of the trace.
In order to do so, it is useful to introduce a dual basis as
\eq{
&
^\vee\langle \vec{i},\vec{j};\alpha| \equiv    \sum_{\{\vec{i'},\vec{j'}\}} (N_{\alpha}^{-1})_{\vec{i}\vec{j},\vec{i'}\vec{j'}} \langle \vec{i'},\vec{j'};\alpha| }
where $N_\alpha^{-1} $ denotes the the matrix inverse of  $N_\alpha$ whose matrix elements are defined in \eqref{inner}.
One can easily verify that the dual basis is indeed orthonormal to the basis \eqref{ketbasis}, such that
\eq{
^\vee\langle \vec{i},\vec{j};\alpha|  \vec{i'}\vec{j'};\alpha\rangle=\delta_{\vec{i},\vec{j}; \vec{i'},\vec{j'}}\delta_{\alpha,\alpha'}.
}
Then the trace in the partition function \eqref{zbth} can be defined by
\begin{equation}\label{torusp1}
Z(a,b)=Tr(e^{-2\pi  a(L_0-\frac{c_L}{24})- 2\pi b(M_0-\frac{c_M}{24})})=\sum_{\vec{i},\vec{j};\alpha}\,^\vee\langle \vec{i},\vec{j};\alpha| e^{-2\pi a(L_0-\frac{c_L}{24})- 2\pi b(M_0-\frac{c_M}{24})} |\vec{i},\vec{j};\alpha\rangle.
\end{equation}
Note that the action of $M_0$ on $|\vec{i},\vec{j},\alpha\rangle$ is to take the eigenvalue of the boost charge, combined with changing one of the $A_{-k}$ to $B_{-k}$, so that
\begin{equation}\label{m0action}
M_0|\vec{i},\vec{j};\alpha\rangle=-{\alpha^2\over2}|\vec{i},\vec{j};\alpha\rangle+\sum_{\{i_k\geq1\}}i_k|i_1,\cdots,i_k-1,\cdots,j_1,\cdots,j_k+1,\cdots;\alpha\rangle.
\end{equation}
From the definition of the dual basis, all terms except the first one have vanishing inner products with the dual state $^\vee\langle\vec{i},\vec{j} ;\alpha |$, and then the expectation value of $M_0$ on this state is just the boost charge of the zero mode
\begin{equation}\label{m0action}
^\vee\langle\vec{i},\vec{j} ;\alpha |M_0|\vec{i},\vec{j};\alpha\rangle=-{\alpha^2\over2}.\end{equation}
The action of $M_0^n$ can be analyzed in a similar way and we learn that the only non-trivial contribution to the expectation value of $e^{- 2\pi a(L_0-\frac{c_M}{24})}$ comes from the zero mode part.
As the zero mode backgrounds $|\alpha\rangle$ all have vanishing conformal weights, the only non-trivial contribution to the operator $e^{- 2\pi a(L_0-\frac{c_M}{24})}$ comes from the non-zero mode part.
Then the torus partition function factorizes into a product of the zero mode part $Z^0(b)$ and the oscillator part $\tilde{Z}(a)$,
\eq{
&Z(a,b)=Z^0(b) \tilde Z(a),\\
&Z^0(b)\equiv \int d\alpha \langle\alpha| e^{- 2\pi b(M_0-\frac{c_M}{24})} |\alpha\rangle=\int_{-\infty}^{\infty}d\alpha e^{2\pi b \alpha^2\over2}+\int_{-i\infty}^{i\infty}d\alpha e^{2\pi b \alpha^2\over2},\nonumber\\
& \tilde Z(a)=\sum_{\vec{i},\vec{j}}\,^\vee\langle \vec{i},\vec{j}| e^{-2\pi a(L_0-\frac{c_L}{24})} |\vec{i},\vec{j}\rangle.\nonumber
}
In the expression of the zero mode part, we have used the fact that $ |\alpha\rangle$ has zero conformal weight and boost charge $\xi=-\frac{\alpha^2}{2}$. Note that the boost charge can be any real numbers, so that we have to integrate along both the real axis and the imaginary axis. The integral can be calculated by analytic continuation from $Re(b)=0$. We get the zero mode contribution,
\begin{equation}\label{znot}
Z^0(b)=\int_{-\infty}^{\infty}d\alpha \left(e^{\pi b \alpha^2}+e^{-\pi b \alpha^2}\right)=\sqrt{\frac{2}{|b|}}.
\end{equation}
Next, we calculate the contribution from the non-zero modes.
Note that either the $A_{-k}$ or the $B_{-k}$ operator raises the weight of $L_0$ by $k$, which enables us to
split the total eigenvalue into a sum of contributions from the $A$ and $B$ modes separately, so that we have
\begin{equation}
\tilde Z(a) \equiv q^{-\frac{1}{12}}\prod_{k=1}^{\infty} \Big(\sum_{i_k=0}^\infty  q^{k \, i_k }\Big)\Big(\sum_{j_k=0}^\infty  q^{k \, j_k }\Big)
=q^{-\frac{1}{12}}\prod_{k=1}^{\infty}\frac{1}{(1-q^k)^2}=
\frac{1}{\eta^2(ia)}
\label{ztilde}
\end{equation}
where
$\eta(ia)$ is the Dedekind-$\eta$ function,
\begin{equation}
\eta(ia)=q^{\frac{1}{24}}\prod_{k=1}^{\infty}(1-q^k),\quad q=e^{-2\pi a }
\end{equation}
Finally, combing the zero mode part \eqref{znot} and the oscillator part \eqref{ztilde} we obtain the torus partition function in the highest weight vacuum,
\begin{equation}\label{tp625h}
Z(a,b)=\sqrt{\frac{2}{|b|}}\frac{1}{\eta^2(ia)}
\end{equation}

As a consistency check, let us now compute how the torus partition function transforms under the modular S transformation,
\begin{equation}
a\rightarrow \frac{1}{a},\ \ \ b\to -\frac{b}{a^2},\ \ \end{equation}
Using the transformation property of the eta function
\eq{\label{de627t}\eta(i\frac{1}{a})=\sqrt{a}\eta(ia)
}
we indeed obtain the relation \eqref{modularinv}, and hence confirm with the general argument \cite{Bagchi:2019unf, Bagchi:2012xr,Barnich:2012xq, Song-Wen-Jiang} that the BMSFT torus partition function is modular invariant.

\subsubsection*{From the UR limit}
In this subsection, we calculate the torus partition function from the UR limit of a free scalar CFT$_2$ in the flipped representation.
To do so, we first need to work out the CFT$_2$ partition function in the flipped vacuum.
Recall that the torus partition function in the highest weight vacuum of the free scalar CFT$_2$ reads,
\begin{equation}\label{t628hpc}
Z^{CFT}_{H}=\frac{q^{-\frac{c}{24}}\bar q^{-\frac{\bar c}{24}}}{\sqrt{Im\tau}}\prod_{k=1}^{\infty}\sum_{n_k=0}^{\infty}q^{kn_k}\bar{q}^{kn_k}=\frac{1}{\sqrt{Im\tau}}\frac{1}{\eta(\tau)\bar{\eta}(\bar \tau)}
\end{equation}
where
\begin{equation}
c=\bar{c}=1,\ \ q=e^{2\pi i \tau},\ \ \bar q=e^{2\pi i\bar\tau}
\end{equation}
One can use \eqref{auto_cft_modes} to map the highest weight vacuum to the flipped vacuum,
 or equivalently
\begin{equation}
\bar \tau\rightarrow -\bar \tau.
\end{equation}
Using the above map, we obtain the torus partition function in the flipped vacuum \footnote{In the discussion below, we only consider the case where $Im\tau>0$ for simplicity. After taking the UR limit, it turns to the case $a>0$. To get the other case $a<0$, one should start with $Im\tau<0$, by a similar consideration. Note that the absolute value of $b$ also follows from this reasoning.},
\begin{equation}
Z^{CFT}_{F}=\frac{q^{-\frac{c}{24}}\bar q^{-\frac{\bar c}{24}}}{\sqrt{Im\tau}}\prod_{k=1}^{\infty}\sum_{n_k=0}^{\infty}q^{kn_k}\bar{q}^{-kn_k}=\frac{1}{\sqrt{Im\tau}}\frac{1}{\eta(\tau)\eta(\bar \tau).}\label{Zf}
\end{equation}
Under the ultra-relativistic limit \eqref{URlim},
the modular parameters become
\eq{\label{mp632ur}\tau=\tilde{a}+ib\epsilon,\,\quad}
To get the result in the Lorentzian theory, we should further take the analytic continuation $\tilde{a}\rightarrow ia$.
Taking the limit $\epsilon\to0$ with this taken into account, we find that the CFT$_2$ in the flipped representation
torus partition function \eqref{Zf}  indeed becomes that of the free scalar BMSFT in the highest weight vacuum \eqref{tp625h}, namely
\begin{equation}
Z_F^{CFT}\rightarrow  \frac{1}{\sqrt{\epsilon}} Z^{BMS}_{H}.
\end{equation}
where the overall factor comes from the rescaling of the field \eqref{model_UR_limit}.

\section*{Acknowledgments}
We would like to thank Luis Apolo, Arjun Bagchi, Bin Chen, Reiko Liu, Wenxin Lai, Zhefei Yu, and Yufan Zheng for useful discussions.
We would like to specially thank Bin Chen and Reiko Liu for helpful discussions on the staggered module, and Luis Apolo for reading the draft thoroughly and for his valuable comments.
WS would like to thank the Okinawa Institute of Science and Technology (OIST) Quantum Gravity group, and workshop on ``String theory and related physics'' where some partial results of this work were presented and helpful comments were received.
The work is partially supported by National Natural Science Foundation of China NO. 11735001, national key research and development program of China NO. 2020YFA0713000, and  Beijing Municipal Natural Science Foundation NO.  Z180003.

\vspace{1cm}
\appendix
\renewcommand{\appendixname}{Appendix~\Alph{section}}

\section{Radial quantization in BMSFT}

In this appendix we provide a prescription of radial quantization in BMSFTs. We first review the procedure in relativistic quantum field theory as reviewed in \cite{Roberts:2014ifa}, and then extend it to BMSFTs.

\subsection{Analytic continuation from Euclidean theory to Lorentzian theory}
Let us start with the Euclidean correlation function of a relativistic quantum field theory.
\begin{equation}
\langle O_1(t^E_1)O_2(t^E_2)\rangle=\langle 0|O_1(0)e^{-H(t^E_1-t^E_2)}O_2(0)|0\rangle\theta(t^E_1-t^E_2)+\langle 0|O_1(0)e^{-H(t^E_2-t^E_1)}O_2(0)|0\rangle\theta(t^E_2-t^E_1)\label{Ecorrelators}
\end{equation}
which is automatically time ordered as $e^{-H\delta t^E}$ is unbounded for $\delta t^E<0$.
To get correlation functions in the Lorentzian theory, one needs to first
analytically continue the time direction to the complex plane, with the Euclidean time as the imaginary part,
\begin{equation}
t'=-it^E+t,
\end{equation}
so that the Lorentzian theory corresponds to the real-time theory with $t^E=0$, and the Euclidean theory corresponds to the imaginary-time theory with $t=0$.
In the correlators \eqref{Ecorrelators}, one can turn on a real part $t$  with the imaginary part $t_E$ fixed,
so that the ordering is still controlled by $t_E$.
As a final step, we need to take the limit $t^E=0$ along a chosen trajectory  \eq{\label{pres}t_E=\lambda \mathcal F(t),\quad \lambda\to0} where $\mathcal F(t)$ is some function of $t$.
Then the ordering in $t_E$ is transferred to $t$.
For example, to get the usual time-ordered correlators in Lorentzian QFT, we can choose\begin{equation}
t^E_i=\lambda t_i, \quad \hbox{with}\quad \lambda\to0.
\end{equation}
so that the ordering in the imaginary time $t_E$ is the same as that of the real time.

\subsection{Radial quantization in CFT$_2$}
Now let us consider a CFT$_2$ on the Euclidean cylinder $(\sigma, t_E)$, with the identification
\begin{equation}
(\sigma,t_E)\sim(\sigma+2\pi,t_E).
\end{equation}
After the analytic continuation $t_E\rightarrow t_E+it$, the complex coordinates become
\eq{
w&=\sigma-it_E\to\sigma+t-it_E,\\
\bar w&=\sigma+it_E\to\sigma-t+it_E.
}
and the cylinder to plane map becomes
\eq{\label{plane2cylinCFT2}
z=e^{iw}=e^{i(\sigma+t)}e^{t_E},\quad
\bar z=e^{-i\bar w}=e^{-i(\sigma-t)}e^{t_E}.
}

After the analytical continuation, $z$ and $\bar z$ are no longer complex conjugate to each other, and effectively we have extended the theory from $\mathbb C$ to $\mathbb C^2$. The Euclidean theory can be obtained by imposing $z^\ast=\bar{z}$,  while the Lorentz theory corresponding to taking $|z|=|\bar z|=1$.
In the Euclidean theories with $t=0$, the infinitely past $t_E\rightarrow-\infty$ on the cylinder becomes the origin on the plane.
Therefore radial quantization at fixed radius on the Euclidean plane corresponds to canonical quantization at fixed time $t_E$ on the cylinder,
As a result,  radial ordering on the Euclidean plane corresponds to Euclidean time ordering on the cylinder, which after a prescription in the form
of \eqref{pres} provides an ordering in the real time $t$.
\subsection{Radial quantization in BMSFT}
Now we provide a prescription of radial quantization and time ordering in BMSFT.
From a CFT$_2$ on the cylinder with complexified time,
we take the UR limit $t \rightarrow\epsilon \tau$, with $t_E$ fixed, so that
\begin{equation}\label{Ccyl}
w=\sigma+\epsilon \tau-it_E,
\end{equation}
\begin{equation}\label{Ccylbar}
\bar w=\sigma-\epsilon \tau+it_E.
\end{equation}
Under the UR limit $\epsilon\to0$, the cylinder to plane map can be defined as
\begin{equation}\label{cplane}
x\equiv \lim\limits_{\epsilon \to 0}\ z=e^{i\sigma+t_E},\ \ y = \lim\limits_{\epsilon \to 0} \frac{z-\bar{z}^{*}}{2\epsilon} \equiv \lim_{\epsilon\to0} {e^{i w}-e^{i\bar{w}^{*}}\over 2\epsilon }=ix\tau
\end{equation}
This is equivalent to the cylinder to plane map \eqref{planeUR} on the unit circle where the latter is defined.

This way we extend the BMSFT to a theory defined on \(\CC \times \R\) where \(\frac{y}{x} \in i\R\) and \(x\) takes arbitrary value on the complex plane \(\CC\), with \(e^{t_{E}}\) as the radius. The pure Euclidean theory corresponds to taking \(\tau = 0\) on the complexified cylinder \eqref{Ccyl} and \eqref{Ccylbar}, or equivalently \(y = 0\) on the complex plane \eqref{cplane}. The Lorentzian theory corresponds to taking \(t_{E} = 0\) in \eqref{Ccyl} and \eqref{Ccylbar}, or equivalently on the \(\left|x\right| = 1\) circle.

In the Euclidean theory, we can perform radial quantization, which again is equivalent to canonical quantization on the cylinder. Correlation functions are radial ordered. Starting from the Euclidean theory on the plane parameterized by $x$, we can obtain the Lorentzian theory by the following steps. First, we fix $t_E$, and turn on $y$ by using the translational generator $M_{-1}$ as in \eqref{transs}, so that the operators and correlators will all depend on the holomorphic coordinate $x$, as well as another complex coordinate $y$. In most part of this paper, we consider the BMSFT on the manifold parameterized by complex coordinates \(x\) and \(y\). Next, we impose the relation $t_E=\lambda \tau$, so that the origin on the Euclidean plane is mapped to past infinity on the Lorentzian  cylinder $\tau=-\infty$. This enables us to establish the operator-state correspondence as in \eqref{operator-state}.
Furthermore, the radial ordering $X(O\cdots O)$ on the Euclidean plane also corresponds to time ordering of the Lorentzian theory.
The final step to get the Lorentzian theory is to take the limit $\lambda\to 0$, so that the theory is restricted on the unit circle. In many discussions, we will not take this final step explicitly.

\section{Comments on the induced representation}
In this section we make some comments on the induced representation.

Since the BMS algebra is the semi-direct sum of the Virasoro algebra and an Abelian ideal generated by $M_n$'s,
one can consider a representation induced from that of the ideal.
In particular, we are interested in the special case when the operators in an indecomposable representation satisfy the following conditions
\eq{
[L_0,O]&=\Delta O,\qquad [M_0,O]=\xi O\nonumber, \\
 [M_n,O]&=0,\,\ \ \quad\quad\, \forall n\neq0, \,n\in \mathbb{Z}.\label{induced}
}
Especially, the induced vacuum denoted as $|0_I\rangle$ should satisfy
\begin{equation}\label{i72v}
L_0|0_I\rangle=M_n|0_I\rangle=0,\ \ \quad\quad \forall n\in \mathbb{Z}.
\end{equation}
As was discussed in \cite{Barnich:2014kra, Barnich:2015uva, Campoleoni:2016vsh}, the induced representation is unitary.
It is also possible to discuss multiplets in the induced representation, where either $L_0$ or $ M_0$ is assumed to be non-diagonalizable, featuring Jordan blocks. We leave this interesting generalization to future study.
In this section, we comment on the induced vacuum for the free BMS scalar model. We find an intrinsically defined vacuum that behaves as a direct product state, whereas another induced vacuum from the UR limit of a relativistic scalar is singular.

\subsection{Induced vacuum from intrinsic discussion}
In section 3.2.1, we discussed how to find a vacuum that is annihilated by $L_{0,\pm1},\,M_{0,\pm1}$, and meanwhile can be described intrinsically by the $A_n$, $B_n$ modes. We note that there are two different choices, one of which is the highest weight vacuum. Now we turn to the other choice. Choosing the condition I and II, we get another vacuum satisfying
\begin{equation}\label{i73v1}
B_n|0_I\rangle=0,\ \ \ \forall n\in\mathbb{Z}.
\end{equation}
This means that all the $A_n$s are creation operators, and all the $B_n$s are annihilation operators. Using these conditions, we learn that the action of $L_n,M_n$ on this vacuum is given by,
\begin{equation}
L_n|0_I\rangle=0,\ \ M_n|0_I\rangle=0,\ \ \ \forall n\in \Z.
\end{equation}
This vacuum satisfies \eqref{i72v}, hence is an induced vacuum.
To study the property of this vacuum, we calculate the Green's function with respect to $x$-ordering on this vacuum as
\begin{equation}
\langle \phi(x_1,y_1)\phi(x_2,y_2)\rangle =-2\pi i \bigg( y_1 \theta(x_1-x_2) +y_2\theta(x_2-x_1) \bigg) \delta(x_2-x_1).
\end{equation}
The correlator above tells us that there is no correlation between two points with different spatial coordinates $x$ on the plane, or $\sigma$ on the cylinder. In other words, this vacuum behaves as a direct product of states living at each point of the spatial slice.

\subsection{Induced vacuum from the UR limit}
Note that the condition \eqref{i73v1} is equivalent to
\begin{equation}
:B_nB_m:|0_I\rangle=0, \ \ \forall n,m \in\mathbb{Z},
\end{equation}
which means that every term in $M_n$ annihilates the vacuum.  There may exist different vacua satisfying \eqref{i72v}, but not the condition above.  The induced vacuum from the UR limit \cite{Bagchi:2017cpu} provides an example of this type.
More explicitly,
taking the UR limit \eqref{URlim} and \eqref{contraction}
of the usual highest weight representation in CFT$_2$, \begin{equation}
L^+_0O=hO,\ \ {L}^-_0O=\bar{h}O,
\end{equation}
\begin{equation}
L^+_{n}O=L^-_{n}O=0,\ \ n>0,
\end{equation} we can get the induced representation of BMS algebra \eqref{induced} with
\begin{equation}\label{h522i}
\Delta=h-\bar{h},\ \ \ \xi=\epsilon(h+\bar{h}),
\end{equation}

\subsubsection*{In the free scalar model}
Now we apply this process to the free scalar. The highest weight vacuum before the UR limit is specified by
\begin{equation}
a_{n}|0_H\rangle_{CFT}=0,\ \ \bar{a}_{n}|0_H\rangle_{CFT}=0, \ \ n>0
\end{equation}
From the relation between modes \eqref{modeslimit}, one might conclude that under the UR limit $\epsilon\to 0$,
the BMS induced vacuum becomes the vacuum specified by \eqref{i73v1}.
However, the UR limit has to be taken more carefully. The reason is as follows.
Consider a basis  constructed by applying $A_n$s and $B_n$s successively on the UR limit of the highest weight vacuum.
By calculating the Gram matrix, we find that none of the basis states become null, and furthermore we cannot find any simple linear combinations of such basis that become null.  For example, $\langle0_H |A_{-n} B_n|0_H\rangle_{CFT}={1\over 2}, \, \forall n\neq0$, so that \eqref{i73v1} is not satisfied, and therefore the UR limit of the highest  weight vacuum is different from the induced vacuum from intrinsic discussion.
In fact, it is unclear if it is possible to express the UR limit of the highest weight vacuum intrinsically in terms of modes $A_n$s and $B_n$s.

Nevertheless, let us try to study some properties of the induced vacuum from its parent CFT$_2$ theory and the limiting procedure.
To calculate the Green's function of $\phi$, we exploit the relation \eqref{modeslimit} and perform the calculation in terms of the modes $a_n,\bar{a}_n$. This is equivalent to taking the UR limit directly from the Green's function in the relativistic free scalar. As a result, we get the Green's function for the BMS scalar $\phi$ on the cylinder \begin{equation}
\langle\phi(\sigma_1,\tau_1)\phi(\sigma_2,\tau_2)\rangle=-\frac{1}{2\epsilon}\log(2-2\cos \sigma_{12})
\end{equation}
where the divergence comes from the rescaling of the field in the limit \eqref{model_UR_limit}.
Similarly, the UR limit of Green's function on the plane
\begin{equation}
\langle\phi(x_1,y_1)\phi(x_2,y_2)\rangle = \frac{1}{\epsilon}\bra \Phi\Phi \ket=-\frac{1}{2\epsilon}(\log (x_1-x_2)+\log(\frac{1}{x_1}-\frac{1}{x_2})).\label{planecor}
\end{equation}
Putting the issue of divergence aside,  the Green's function on the plane has the property that it depends on both complex coordinates  $x_1$ and $x_2$, instead of the difference $x_1-x_2$.
This is because the induced vacuum \eqref{induced} is not necessarily translationally invariant on the $x$ plane. The divergence of the Green's function on both the cylinder and the plane reflects the difficulty in finding this induced vacuum intrinsically in terms of annihilation creation operators of our BMSFT model. The divergence, however, is potentially related to the divergence in the one-loop partition of three dimensional gravity discussed in \cite{Barnich:2015mui}.

\subsubsection*{Torus partition function}
We will consider the torus partition function \eqref{zbth} of BMS free scalar in the induced vacuum from the UR limit. Since we cannot deal with it intrinsically using the $A_n,B_n$ modes, it is convenient to express the state space in an orthonormal basis of the CFT$_2$,
\begin{equation}
|\vec{i},\vec{j};\alpha\rangle_{CFT}=a_{-1}^{i_1}\cdots\bar a_{-1}^{j_1}\cdots|\alpha\rangle_{CFT}
\end{equation}
where $|\alpha\rangle_{CFT}$ is the zero mode contribution in the CFT$_2$ highest weight vacuum. The behaviour under $L_0,M_0$ can be calculated from \eqref{contraction}, \begin{equation}
L_0|\alpha\rangle_{CFT}=0,\ \ M_0|\alpha\rangle_{CFT}=-\frac{\alpha^2}{2}
\end{equation}
Moreover from \eqref{contraction}, the action of the BMS generators $L_0,M_0$ on these states are
\eq{
[L_0,a_m]&=-ma_{m},\ \ [L_0,\bar a_m]=-m\bar a_{m}\\
[M_0,a_m]&= [M_0,\bar a_m]=0
}
Then a similar calculation as in section 6 leads to the torus partition function on the induced vacuum,
\begin{equation}\label{tp631i}
Z=\sqrt{\frac{2}{|b|}}\prod_{k=1}^{\infty} \Big(\sum_{i_k=0}^\infty  q^{k \, i_k }\Big)\Big(\sum_{j_k=0}^\infty  q^{-k \, j_k }\Big)
\end{equation}
The summation in either the first or the second parenthesis becomes divergent as long as $a$ is real.
The similar divergence appears in the calculations of the character of the BMS induced module \cite{Oblak:2015sea} and the one-loop partition function of the asymptotic flat Einstein gravity in three dimensional spacetime \cite{Barnich:2015mui}, where a imaginary part of $a$ is introduced as a regulator to do the summation.

We end this section by the following comments,
\begin{itemize}
  \item The BMS algebra is the semi-direct sum of the Virasoro algebra and an Abelian ideal generated by the $M_n$s, so
one can consider a representation induced from that of the ideal. We are interested in a special kind of induced representation defined by \eqref{induced},  with the vacuum specified by the condition \eqref{i72v}. Note that there may be more than one vacua satisfying this condition \eqref{i72v}.
  \item From the mode expansion of our free BMS scalar model \eqref{mode}, one can define an induced vacuum intrinsically \eqref{i73v1} obeying \eqref{i72v}, which behaves as a direct product state.
  \item Another vacuum satisfying \eqref{i72v} comes from the UR limit of free scalar model, as the original general discussion. This vacuum cannot be expressed intrinsically in terms of the modes $A_n,B_n$. It leads to a different theory whose Green's function and torus partition function are both divergent as $\epsilon\rightarrow0$.
\end{itemize}

\section{NR limit from CFT$_2$}
Here we review the non-relativistic (NR) limit of CFT$_2$ for completeness. We will see that the NR limit on the plane is quite different from the UR limit on the plane.
\subsection*{NR limit on the cylinder}
Starting with the same setup as in the section 5.1.1, the NR limit is
 \begin{equation}\label{NRlim}
\sigma =\epsilon s,\ \quad \epsilon\rightarrow 0 \end{equation}
so that the speed of light goes to infinity, which is the reason why this limit is called the non-relativistic limit. The theory is defined on $(s,t)$ after the NR limit. Compared with the UR limit on the cylinder \eqref{URlim}, the NR limit rescales the $\sigma$ direction instead of the $t$ direction. Under this limit the conformal transformations become the GCA transformations generated by
\begin{equation}
l_n=t_n+\bar{t}_{n}=ie^{in\tau}(\partial_\tau+in\sigma\partial_\sigma),\ \ \ m_n=\epsilon(t_n-\bar{t}_n)=ie^{in\tau}\partial_\sigma
\end{equation}
The Virasoro algebra \eqref{CFT2alg} becomes the GCA algebra, which is isomorphic to \eqref{BMSalg} via another Wigner-In\"{o}n\"{u} contraction \cite{InonuWigner},
\begin{equation}\label{NRB2c}
L_n=T_n+\bar{T}_n,\ \ M_n=\epsilon(T_n-\bar{T}_n)
\end{equation}
with the central charges related by,
\begin{equation}\label{NRc}
c_L=c+\bar{c},\ \ c_M=\epsilon(c-\bar{c}).
\end{equation}
\subsection*{NR limit on the plane}
We use the map \eqref{cft_map} to go to the plane $(z,\bar{z})$ before the NR limit, with the CFT$_2$ generators given by \eqref{c511g}. Now we can take the plane NR limit \eqref{naiveplane} safely to get the GCA generators,
\begin{equation}
l_n=t_n+\bar{t}_{n}=-x^{n+1}\partial_x-(n+1)x^ny\partial_y,\ \ \ m_n=\epsilon(t_n-\bar{t}_n)=-x^{n+1}\partial_y,
\end{equation}
which is consistent with the GCA cylinder-to-plane map
\begin{equation}x=e^{i\tau},\ \ y=ie^{i\tau}\sigma.
\end{equation}

\subsubsection*{Representation}
We can consider the usual highest weight representation in CFT$_2$,
\begin{equation}
L^+_0O=hO,\ \ {L}^-_0O=\bar{h}O,
\end{equation}
\begin{equation}
L^+_{n}O=L^-_{n}O=0,\ \ n>0.
\end{equation}
Under the NR limit \eqref{NRB2c}, the highest weight representation becomes the highest weight representation of the GCA algebra,
\begin{equation}
L_0O=\Delta O,\ \ M_0O=\xi O,
\end{equation}
\begin{equation}
L_{n}O=M_{n}O=0,\ \ n>0,
\end{equation}
where
\begin{equation}\label{c11nr}
\Delta=h+\bar{h},\ \ \ \xi=\epsilon(h-\bar{h}).
\end{equation}

The flipped representation in CFT$_2$ satisfies
\begin{equation}
L^+_0O=hO,\ \ L^-_0O=-\bar{h}O,
\end{equation}
\begin{equation}
L^+_{n}O=L^-_{-n}O=0,\ \ n>0.
\end{equation}
Under the NR limit \eqref{NRB2c}, one gets the induced representation of the GCA algebra.
\begin{equation}
L_0O=\Delta O,\ \ M_0O=\xi O,
\end{equation}
\begin{equation}
M_{n}O=0,\ \ n\neq 0,
\end{equation}
where
\begin{equation}\label{h522i}
\Delta=h-\bar{h},\ \ \ \xi=\epsilon(h+\bar{h}).
\end{equation}

\subsubsection*{The free scalar model}
At the level of representation, the UR limit of the flipped representation shares the same features as those of the NR limit of the highest weight representation. However,  our BMSFT model \eqref{action} is different from the NR limit of a CFT$_2$.

Starting from a free scalar model in the relativistic theory \eqref{rel_scalar_action}, the NR limit \eqref{NRlim} along with the field rescaling $\Phi=\sqrt{\epsilon}\phi$ leads to a non-relativistic theory with action
\eq{ S=\frac{1}{4\pi}\int dsd\tau(-(\partial_s\phi)^2)}
Note that the above action is similar to but different from that of our BMSFT model \eqref{action}. First,  the kinematic term has the wrong sign, and a further Wick rotation of the field $\phi\rightarrow i\phi$ is needed to fix the sign. Second, the derivative in the kinematic term is in different directions from the BMS scalar action \eqref{action}. As we perform canonical quantization along the spatial circle, the canonical quantization of the the two limits will be different.
Third, after the NR limit, the spatial cycle becomes noncompact, whereas in the UR limit the circle remains compact.

\bibliographystyle{JHEP}
\bibliography{BMS scalar,refs}

\end{document}